%% file: main.tex
\documentclass{llncs}
\usepackage{makeidx}  

\input{Commands}

\newcommand{\TheTitle}{Temporal logic control of general Markov decision processes by approximate policy refinement}

\pgfplotsset{compat=1.14}
\begin{document}
\frontmatter          
\pagestyle{headings}  
\addtocmark{Symbolic control of general Markov decision processes by approximate stochastic policy refinements} 

\mainmatter              
\title{\TheTitle}
\titlerunning{Policy refinement via approximate similarity relations}  
	
\author{Sofie Haesaert\inst{1} \and
Sadegh Soudjani\inst{2} \and
Alessandro Abate\inst{3}}
\authorrunning{Sofie Haesaert et al.} 
%
%
\institute{California Institute of Technology, United States
\and
School of Computing, Newcastle University, United Kingdom
\and
Computer Science Department, Oxford University, United Kingdom}

\maketitle              

\begin{abstract}
The formal verification and controller synthesis for Markov decision processes that evolve over uncountable state spaces are computationally hard and thus generally rely on the use of approximations. In this work, we consider the correct-by-design control of general Markov decision processes (gMDPs) with respect to temporal logic properties by leveraging  approximate probabilistic relations between the original model and its abstraction. 
We newly work with a robust satisfaction for the construction and verification of control strategies, which allows for  both deviations in the outputs of the gMDPs and in the probabilistic transitions.
 The computation is done over the reduced or abstracted models, such that when a property is robustly satisfied on the abstract model, it is also satisfied on the original model with respect to a refined control strategy.
\end{abstract}

\section{Introduction}
With the ever more ubiquitous embedding of digital components into physical systems 
new computational efficient verification and control synthesis methods for these cyber-physical systems  are needed.
The correct functioning of cyber-physical systems can only be expressed over the combined behaviour of both the digital component and its connected physical system. Quite importantly, stochastic models are key when computers interact with physical systems such as biological processes, power networks, and smart-grids.
These dynamic systems with uncertainty and non-determinism can be modelled as Markov processes evolving over continuous spaces.
Their potential safety critical impact on the environment they interact with makes it of particular interest to develop formal methods that assist in their verifiable design. 
In this work, we newly enable the verification and synthesis of these stochastic systems with respect to probabilistic linear temporal logic properties.

As our modelling framework, we consider the rich class of general Markov decisions processes (gMDPs), which are Markov decision processes evolving over continuous or uncountable state spaces and which have control-dependent stochastic transitions in combination with a metric output space. It is over this output that we define properties of interest.
The characterisation of properties over such processes can in general not be attained analytically \cite{Abate1}, 
so an alternative is to approximate these models by simpler processes, such as finite-state MDP \cite{SAID} or continuous-space reduced order models \cite{safonov1989schur} that are prone to be mathematically analysed or algorithmically verified \cite{FAUST13}.
In \cite{haesaert2017verification,DBLP:conf/qest/HaesaertAH16} we have proposed new approximate similarity relations to quantify the accuracy of the approximation utilising bounds on the output distance and on the transition probabilities. We have shown that for bounded safety properties these approximate similarity relations can be used to refine control strategies with bounded error on transition probability and deviations in the output space. 
The main goal of this paper is to study general temporal logic properties of gMDPs in combination with these approximate similarity relations. As the main contribution of the paper we show that the standard verification and control synthesis for gMDPs can be made robust a-priori to the introduced accuracy errors. 

\smallskip

\noindent\textbf{Related Work.}
Properties defined in PCTL, PLTL, and PCTL$^*$ for finite-state Markov (decision) processes can be verified using tools such as PRISM \cite{KNP11}.
Moreover it is also well-known how to design policies, i.e., to control these Markov decision processes such that the satisfaction of these properties is maximised.
The work in \cite{AbateQuanti} has studied model-checking of automata specifications against autonomous (i.e. uncontrolled) discrete-time stochastic models over uncountable state spaces. 
It was shown that the computation of the probability of satisfying a specification expressed as deterministic finite automaton (DFA) can be restated in terms of a probabilistic reachability problem over the product between the original model and the DFA.
This result has been extended in \cite{tmka2013} to the case of controlled discrete-time Markov processes, which are a special subclass of gMDPs introduced in our work (obtained with identity output map, $h(x)=x$ for all $x\in\X$, and so with an output space $\Y=\X$).
In this work we extend the model class discussed in \cite{tmka2013} and build on their results by showing that an approximate model can be used to compute the solutions of Bellman equations associated to the  stochastic reachability problem. 

\medskip

The paper is organised as follows. In the next section, we first define gMDPs and state the temporal logic control problem. In Section \ref{sec:probreach}, we define approximate simulation relation on gMDPs and solve the robust probabilistic reachability problem. In Section \ref{sec:fullcase}, we extend the results to probabilistic temporal logic control utilising the product of gMDP and DFA. 
In Section \ref{sec:case_study}, we detail the approximation procedure for linear stochastic dynamical systems and we apply it to a simple toy case. Throughout the paper proofs of the theorems have been relegated to the appendix of the extended version \cite{tech_report_TACAS}.

\input{ProblemsStatement}
\section{Quantifying stochastic reachability using $(\eps,\delta)$-probabilistic simulation relations} \label{sec:probreach}

Let a gMDP $\M=(\X,\pi,\mathbb T,\A,h,\Y)$ and a target set $K\subset \Y$ be given.
In the robust formulation of Problem~1, we compute a control strategy $\Ca$ and a quantified lower bound $p$, such that  the probability of satisfying $\eventually^{\leq N} K $, respectively $\eventually K $,
is lower bounded by $p$,
i.e.,
$
\pcm{\Ca}{\M}(\lozenge^{\le N} K) \geq p \mbox{, respectively }  
\pcm{\Ca}{\M}(\lozenge K)\geq p.
$
We associate to the target set $K\subset \Y$, the corresponding set in the state space  $K_\X:= h^{-1}(K)\in \mathcal B(\X)$. Let us denote by $r^\mu(K_\X,N)$,
the probability that $K$ is reached within $N$ time steps when a Markov policy $\mu=(\mu_0,\mu_1,\ldots, \mu_{N-1})$ is used.
We iterate the computation of stochastic reachability of a Markov decision process, as explain in \cite{Abate1}.
The value of  $r^\mu(K_\X,N)$ can be computed by a backward recursion
initialised with $ V_N =0$, and iterated for $ k=N-1,\ldots,0$ as
\begin{equation}
\label{eq:V_rec}
V_k(x) = \int_{\X}\left[\mathbf 1_{K_\X}(\bar x)+ \mathbf 1_{\X\setminus K_\X}(\bar x) V_{k+1}(\bar x)\right]\mathbb T(d\bar x|x,\mu_k(x)).
\end{equation}
Based on the final value function after $N$ iterations, we have that
\begin{equation}
\label{eq:prob_reach_finite}
r^\mu(K_\X,N) = \int_\X\left[ \mathbf 1_{K_\X}(x)+ \mathbf 1_{\X\setminus K_\X}(x)V_0(x)\right] \pi(dx).
\end{equation}
Furthermore,
the optimal value functions $V_k^*(x)$, $k\in[0,N]$, computed recursively with $V_N^*(x) = 0$ and for all $x\in\X$ and $k=N-1,\ldots, 1,0$,
\begin{equation}
\label{eq:V_recopt}
V_k^*(x) = \sup_{\mu_k}\int_{\X}\left[\mathbf 1_{K_\X}(\bar x)+ \mathbf 1_{\X\setminus K_\X}(\bar x) V^*_{k+1}(\bar x)\right]\mathbb T(d\bar x|x,\mu_k(x)),
\end{equation}
give the optimal reachability probability
\begin{equation}
\label{eq:prob_reach_inf}
r^*(K_\X,N) = \int_\X\left[ \mathbf 1_{K_\X}(x)+ \mathbf 1_{\X\setminus K_\X}(x)V_0^\ast(x)\right] \pi(dx).
\end{equation}
Using $V_k^*(x)$ the control strategy $\Ca^*$ maximising $\pcm{\Ca}{\M}(\eventually K)$ is defined by a Markov policy  $\mu^*$ with elements $\mu_k^*$
\begin{equation}
\label{eq:policy_rec}
\mu_k^\ast(x) \in\arg\sup_{\mu_k} \int_{\X}\left[\mathbf 1_{K_\X}(\bar x)+ \mathbf 1_{\X\setminus K_\X}(\bar x) V_{k+1}^*(\bar x)\right]\mathbb T(d\bar x|x,\mu_k(x)). 
\end{equation}

The unbounded optimal reachability probability  $r^\ast(K_\X)$ can be evaluated based on the fixed point of \eqref{eq:V_recopt}.
More specifically, for $N\rightarrow \infty$ the value functions are strictly increasing and converging to the fixed point solution of 
\begin{equation}
\label{eq:V_recopt_inf}
V^*(x) = \sup_{\mu}\int_{\X}\left[\mathbf 1_{K_\X}(\bar x)+ \mathbf 1_{\X\setminus K_\X}(\bar x) V^*(\bar x)\right]\mathbb T(d\bar x|x,\mu(x)).
\end{equation}
For a given policy $\mu$, the unbounded reachability probability $r^\mu(K_\X)$ and the associated value function $V^\mu(x)$ are formulated similarly.
Thus we have that
\begin{equation*}
\sup_\Ca \pcm{\Ca}{\M}(\lozenge K)=r^*(K_\X) \mbox{, respectively, }  \sup_\Ca\pcm{\Ca}{\M}(\lozenge^{\le N} K) = r^*(K_\X,N).
\end{equation*} 
This together, with its universal measurability, has been discussed in \cite{Abate1}. Computation of backward recursions \eqref{eq:V_rec} and \eqref{eq:V_recopt} is generally only tractable for finite state spaces.
In the following, we define approximate probabilistic simulation relations over $\mathcal{M}_\Y$, as introduced in \cite{haesaert2017verification}, and apply it to compute a lower bound on probabilistic reachability and the corresponding synthesis problem. 

\subsection{Approximate simulation relations for gMDPs}
Consider two gMDPs $\M_i=(\X_i,\pi_i,\mathbb T_i,\A_i,h_i,\Y)$, $i=1,2$, that have the same metric output space $\left(\Y,\mathbf d_\Y\right)$.
Given state-action pairs $x_1\in \X_1,u_1\in \A_1$ and $x_2\in \X_2,u_2\in \A_2$, we want to relate the corresponding transition kernels,
namely the probability measures
$\mathbb T_1(\cdot\mid x_1,u_1)\in\mathcal{P}(\X_1,\mathcal{B}(\X_1))$
and $\mathbb T_2(\cdot\mid x_2,u_2)\in\mathcal{P}(\X_2,\mathcal{B}(\X_2))$.
As in \cite{haesaert2017verification}, we introduce the concept of $\delta$-lifting as follows.
\begin{definition}[$\delta$-lifting for general state spaces]
\label{def:del_lifting}
	Let $\X_1,\X_2$ be two sets with associated measurable spaces $(\X_1,\mathcal B(\X_1)), (\X_2,\mathcal B(\X_2))$,
	and let  $\rel\subseteq \X_1\times \X_2$ be a relation for which $\rel\in \mathcal B(\X_1\times \X_2)$.
	We denote by
	$\bar\rel_\delta\subseteq \mathcal{P}(\X_1,\mathcal B(\X_1))\times \mathcal{P}(\X_2,\mathcal B(\X_2))$ the corresponding lifted relation so that $\Delta \bar \rel_\delta \Theta$ holds if there exists a probability space $(\X_1\times \X_2,\mathcal B(\X_1\times \X_2), \mathbb W)$  (equivalently, a lifting $\mathbb W$) satisfying { \setlength{\parskip}{-1pt}\setlength{\parsep}{0pt}
		\begin{description}
			\item[\textbf{L1.}] for all $X_1\in \mathcal{B}(\X_1)$: $\mathbb W(X_1\times \X_2)=\Delta(X_1)$;
			\item [\textbf{L2.}] for all $X_2\in \mathcal{B}(\X_2)$:  $\mathbb W(\X_1\times X_2)=\Theta(X_2)$;
			\item[\textbf{L3.}] for the probability space  $(\X_1\times \X_2,\mathcal B(\X_1\times \X_2), \mathbb W)$ it holds that
			$x_1\rel x_2$ with probability at least $1-\delta$, or equivalently that $\mathbb{W}\left(\rel\right)\geq1-\delta$.
	\end{description}}%
\end{definition}

We introduce a notion  of approximate probabilistic simulation relations which naturally leads to control refinement. For this, we build on the notion of \emph{interface function} \cite{Girard2009} to define probabilistic simulation relations that allow for
the hierarchical control refinement of two gMDPs
\begin{equation*}
\InF: \A_1\times \X_1\times\X_2 \rightarrow \mathcal{P}(\A_2,\mathcal B(\A_2)).
\end{equation*}
This interface function  $\InF$  is required  
to be a Borel measurable function.
Intuitively, an interface function implements (or refines) any control action synthesised over the abstract model to an action for the concrete model.

\begin{definition}[$(\eps,\delta)$-probabilistic simulation relation]
\label{def:apbsim}
	Consider two gMDPs $\M_i=(\X_i,\pi_i ,\mathbb T_i,\A_i,h_i,\Y),$ $i =1,2$,  over a shared {metric} output space  $(\Y,\mathbf{d}_\Y)$.
	$\M_1$ is $(\epsilon,\delta)$-stochastically simulated by $\M_2$ if there exists an interface function $\InF$ and
	a relation $\rel\subseteq \X_1\times \X_2$, for which there exists a Borel measurable stochastic kernel $\Wt(\,\cdot\,{\mid} u_1,x_1,x_2)$ on $\X_1\times\X_2$ given $\A_1\times\X_1\times\X_2$,
	such that:
	{ \setlength{\parskip}{-2pt}\setlength{\parsep}{-1pt}
		\begin{description}
			\item[\textbf{APS1.}] $\forall (x_1,x_2)\in \rel$, $ \mathbf{d}_\Y\left(h_1(x_1),h_2(x_2)\right)\leq \epsilon$;
			\item[\textbf{APS2.}] $\forall (x_1,x_2)\in \rel$, $\forall u_1\in\A_1$:
			\(\mathbb T_1(\cdot| x_1, u_1)\ \bar \rel_\delta \  \mathbb T_2(\cdot| x_2, \InF(u_1,x_1,x_2)),\) with lifted probability measure $\Wt(\,\cdot\,{\mid}u_1,x_1,x_2)$;
			\item[\textbf{APS3.}] $\pi_1\bar \rel_\delta \pi_2$.
	\end{description} }
	\noindent The simulation relation is denoted as $\M_1\preceq^{\delta}_\eps\M_2$.
\end{definition}
This definition extends the known exact notions of probabilistic simulation in \cite{larsen1991bisimulation},
and the approximate notions of \cite{Desharnais2008,cDAK12} to gMDPs over Polish spaces as elaborated in \cite{haesaert2017verification}.
The Borel measurability for both $\InF$ (see above) and $\Wt$ (as in this definition),
which is used to prove the well-posedness of the controller refinement,
can be relaxed to universal measurability  \cite{haesaert2017verification}.

\subsection{$\delta$-Robust probabilistic reachability}

\begin{definition}[($\eps,\delta$)-Robust satisfaction]
Consider any $\M_1\in\mathcal M_\Y$. We say that a Markov policy $\mu$ for $\M_1$ $(\eps,\delta)$-robustly satisfies $\lozenge^{\le N} K$ with probability $p$ if for every $\M_2\in\mathcal M_\Y$ with $\M_1\preceq^\delta_\epsilon \M_2$  there exists a controller $\Ca_2$ for  $\M_2$ such that $\pcm{\Ca_2}{\M_2}(\lozenge^{\le N} K)\geq p$. 
\end{definition}

For a given universally measurable map $\nu:\X_1\rightarrow\mathcal P(\A_1,\mathcal B(\A_1))$ and constant $\delta$, define the operator $\mathbf T_\delta^\nu:\mathcal F\rightarrow\mathcal F$  acting on the set of functions 
$\mathcal F:=\left\{f:\X_1\rightarrow [0,1] \right\}$
as
\begin{equation}
\label{eq:Toperator}
\mathbf T_\delta^\nu(V)(x):=\Lim \left(\int_{ \X_1}
[\mathbf 1_{K_{\X_1}}( \bar x) +  \mathbf 1_{ \X_1 \setminus K_{\X_1}}( \bar x)  V(\bar x)]
\mathbb T(d  \bar x|x,\nu(x)) \ -\delta\right)
\end{equation}  
with $\Lim:\mathbb R\rightarrow [0,1]$ being the truncation function  $\Lim(\cdot):=\min(1,\max(0,\cdot))$.
Define also the operator $\mathbf T_\delta^{*}(V)$ on $\mathcal F$ as $\mathbf T_\delta^{*}(V)(x):= \sup_\nu\mathbf T_\delta^{\nu}(V)(x)$.

\begin{theorem}\label{thm:delreach}
	A target set $K\subset \Y$ of gMDP $\M_1$ is reached $(0,\delta)$-robustly with Markov policy $\mu$ and with probability  $ r^\mu(K_{\X_1},N)$, where
	\begin{equation}
	\label{eq:prob1}
	r^\mu(K_{\X_1},N):= \Lim\left(\int_{\X_1}[ \mathbf 1_{K_{\X_1}}(x) +\mathbf 1_{ \X_1 \setminus K_{\X_1}}(x)V_0^\delta(x)] \pi_1(d x)-\delta\right),
	\end{equation}
	and $V_0^\delta(x)$ is computed recursively according to 
	$V^\delta_{k}:=\mathbf T_\delta^{\mu_k}(V^\delta_{k+1})$, for $k=N-1,\ldots, 0$
	with initial value function $V_N^\delta=0$.
	If $V_0^{\delta,*}$ is computed similarly as the solution of recursion $V_k^{\delta,*}=\mathbf T_\delta^\ast(V_{k+1}^{\delta,*})$ with initial value function $V_N^{\delta,*}=0$ and $\mu_k^*\in\arg\sup_{\mu_k} \mathbf T_\delta^{\mu_k}(V_{k+1}^{\delta,*})$ then we call $\mu^* = \{\mu_0,\mu_1,\ldots\}$ the optimal \mbox{$(0,\delta)$-robust} policy.
\end{theorem}
Notice that for $\delta=0$ the computation of value functions $V^\delta_{k}$ in Theorem~\ref{thm:delreach} is the same as \eqref{eq:V_rec}.
The proof of Theorem~\ref{thm:delreach}
builds  on the construction of a refined control strategy as has been explained in \cite{haesaert2017verification}. 
For any $\M_2$ such that $\M_1\preceq_0^\delta\M_2$  with the lifted probability measure $\mathbb W_{\mathbb T}$,  the  control policy $\Ca_2$ can be refined from $\Ca_1$ (cf.
\cite{haesaert2017verification}).
More precisely, a control strategy $\mathbf{C}_2$ that refines $\Ca_1$ over $\M_2$ is obtained by extending $\Ca_1$ with internal states $(x_1,x_2)$. 
While the state $(x_1,x_2)$ of $\Ca_2$ is in $\rel$,
the control refinement
has as its \emph{basic ingredients} the states $x_1$ and $x_2$,
whose 
the stochastic transition to the pair $(x_1',x_2')$ is governed firstly by a point distribution $\delta_{x_2(t)}(dx_2')$  based on the measured state $x_2(t)$ of $\M_2$; and subsequently, by the  lifted probability measure
\(
\Wt(dx_1'| x_2', u_1,x_2,x_1),
\) 
\emph{conditioned} on $x_2'$.

Before tackling unbounded reachability properties, we first analyse the behaviour of $\mathbf T_\delta^\nu$ and $\mathbf T_\delta^*$.
Suppose that $W_1(x)\geq W_2(x)$ for all $x\in \X_1$, then for a given map $\nu: \X_1\rightarrow\mathcal P(\A_1,\mathcal B(\A_1))$ we have  
$\mathbf T_\delta^\nu(W_1)(x)\geq \mathbf T_\delta^\nu(W_2)(x),$
hence $\mathbf T_\delta^{*}(W_1)(x)\geq \mathbf T_\delta^{*}(W_2)(x).$
Then the series $V_{N}^\delta,V_{N-1}^\delta,\ldots,V_0^\delta$ constructed with  $V_k^\delta=\mathbf T^{\mu_k}_\delta V^\delta_{k+1}$ and $V_N^\delta=0$, is monotonically increasing.
Therefore, for a given policy $\mu$ the functions $\{(\mathbf T^{\mu_k}_\delta)^{q}(V)\}_{q\ge 0}$ initialised with  $V=0$ is point-wise converging
since it is monotonically increasing and upper bounded.
Additionally, the same holds for the function $ \{(\mathbf T^\ast_\delta)^{q}(V)\}_{q\ge 0}$  initialised with  $V=0$. 
For unbounded  reachability, this yields the following result.
\begin{cor}
\label{thm:delreach_infty}
	A target set $K\subset \Y$ of gMDP $\M_1$ is reached $(0,\delta)$-robustly with time-homogeneous Markov policy $\mu$ and with probability $r^\mu(K_{\X_1})$, where
	\begin{equation}
	    \label{eq:prob2}
	    r^\mu(K_{\X_1}):= \Lim\left(\int_{\X_1}[ \mathbf 1_{K_{\X_1}}(x) +\mathbf 1_{ \X_1 \setminus K_{\X_1}}(x)V^\delta(x)] \pi_1(d x)-\delta\right ),
	\end{equation}
	with $V^\delta:\X_1\rightarrow[0,1]$ the  
	solution of $V^\delta=\mathbf T_\delta^\mu(V^\delta)$ computed as the limit of the sequence $\{(\mathbf T^\mu_\delta)^{q}(V)\}_{q\ge 0}$ that is initialised with $V=0$.
	If $V^{\delta,*}$ is computed similarly as the solution of $V^{\delta,*}=\mathbf T_\delta^\ast(V^{\delta,*})$ and $\mu^*\in\arg\sup \mathbf T_\delta^\mu(V^{\delta,*})$ then we call $\mu^*$ the optimal $(0,\delta)$-robust policy.
\end{cor}

\subsection{$(\epsilon,\delta)$-Robust probabilistic reachability}
%
%
Consider a target set $K\subset \Y$, let $K^\eps$ be the largest Borel measurable set such that 
\begin{equation}
\label{eq:K_erosion}
K^\eps\subset\{y\mid \forall \bar y \in \Y \mbox{ with } \mathbf d_\Y(\bar y,y)\leq\eps:\,  \bar y \in K\}.
\end{equation}
We can now introduce an eroded version of the original target set $K_{\X_1}$ as $K^\eps_{\X_1}:= h_{1}^{-1}(K^\eps), $
such that for any pair $x_1,x_2$ if  $x_1\in K^\eps_{\X_1}$ and $x_1\rel x_2$ then $h_{2}(x_2)\in K$.
As a consequence of Theorem \ref{thm:delreach} and of Corollary \ref{thm:delreach_infty}, we can  evaluate the \mbox{$(\eps,\delta)$-robust} reachability with respect to $K^\eps_{\X_1}$.
\begin{cor}[$(\eps,\delta)$-robust probabilistic reachability]
\label{thm:deleps_infty}
	A target set $K\subset \Y$ is reached $(\eps,\delta)$-robustly with Markov policy $\mu$ for $\M_1$ with respect to $r^{(\delta,\eps)}(K_{\X_1},N)$ (or $r^{(\delta,\eps)}(K_{\X_1})$) if it reaches the target set $K^\eps$ as in \eqref{eq:K_erosion}, $(0,\delta)$-robustly with probability $r^\mu(K_{\X_1}^\eps,N)$ defined in \eqref{eq:prob1} (or $r^\mu(K_{\X_1}^\eps)$ defined in \eqref{eq:prob2}).
	
\end{cor}
Hence for any model $\M_2$ that is in an  $(\eps,\delta)$-probabilistic simulation relation with $\M_1$, 
the combination of Theorem \ref{thm:delreach} and Corollaries \ref{thm:delreach_infty} and \ref{thm:deleps_infty}, mean that we can verify probabilistic reachability robustly over $\M_1$, and moreover, that we can synthesise a robust controller maximising the satisfaction probability robustly.

\subsection{Upper-bounding the probabilistic reachability}
We now question whether we can quantify an upper bound on a reachability probability using an approximate model $\M_1$.
Consider the operator
$\mathbf T_{-\delta}^\mu $, defined as
\begin{equation} 
\mathbf T^\mu_{-\delta}(V)(x):=\Lim \left(\int_{ \X_1}
[\mathbf 1_{K_{\X_1}^{-\eps}}( \bar x) +  \mathbf 1_{ \X_1 \setminus K_{\X_1}^{-\eps}}( \bar x)  V(\bar x)]
\mathbb T(d  \bar x| x, \mu( x)) \ +\delta\right)
\end{equation}
with $ K_{\X_1}^{-\eps}:=h_1^{-1}\left(K^{-\eps}\right)$ and $K^{-\eps}:= \{y+y_\eps\in\Y|y\in K \mbox{ and } \mathbf d_\Y(0,y_\eps)\leq \eps\}$. 
\begin{theorem}\label{thm:maxprob}
	Consider a target set $K\subset \Y$, then an upper bound on the maximal reachability $r^\mu(K_{\X_2},N)$ of $\M_2$ for $\M_1\succeq_{\eps}^\delta \M_2$ can be given as  	\begin{equation}
	\forall \mu:\  r^\mu(K_{\X_2},N)\leq r^{(-\delta,-\eps)}(K_{\X_1},N)
	\end{equation}
	for which $ r^{(-\delta,-\eps)}(K_{\X_1},N):= r^{(-\delta)}(K_{\X_1}^{-\eps} ,N)$ is computed with $\M_1$ as follows,
	\begin{equation*}
	r^{(-\delta,-\eps)}(K_{\X_1},N):= \Lim\left( \int_{\X_1}[ \mathbf 1_{K_{\X_1}^{-\eps}}(x) +\mathbf 1_{ \X_1 \setminus K_{\X_1}^{-\eps}}(x)V_0^\delta(x)] \pi_1(d x) +\delta\right)
	\end{equation*}
	with $V^\delta_k:\X_1\rightarrow[0,1]$ such that $V^\delta_{k}=\sup_\mu\mathbf T_{-\delta}^\mu(V^\delta_{k+1})$ and $V_N^\delta = 0$. 
\end{theorem}
\section{Temporal logic control leveraging $(\eps,\delta)$-probabilistic simulation relations}\label{sec:fullcase}

\input{DFA.tex}
\label{sec:prod}
In the following we analyse the robust satisfaction of scLTL specifications, which are temporal specifications that go beyond reachability properties defined on $\M$. 
The probabily of satisfying such a temporal specification can be quantified as a reachability probability with respect to $\M\otimes\mathcal A_\psi$.  For two gMDPs $\M_1$ and  $\M_2$, subject to $\M_1\preceq_0^\delta\M_2$, we show that ($\delta$-approximate) probabilistic simulation relations are preserved under a product with a DFA.
\begin{theorem}
\label{thm:DFA_product}
	Let $\M_i, i=1,2$, $\M_i = (\X_i,\pi_i ,\mathbb T_i,\A_i,h_i,\Y)$, be two gMDPs such that $\M_1\preceq_0^\delta \M_2$ and $\mathcal A = (Q,q_0,\Sigma,F,\trans)$ be an automaton. For any labeling function $\mathsf L:\Y\rightarrow\Sigma$ we have $\M_1\otimes\mathcal A\preceq_0^\delta \M_2\otimes\mathcal A$.
\end{theorem}
This theorem enables us to quantify temporal logic properties for  $\M_2$ with respect to $\M_1$. 
Consider  a scLTL property $\psi$ with a corresponding DFA $\mathcal A_\psi$ and two gMDPs $\M_1,\M_2\in\mathcal M_\Y$ for which $\M_1\preceq_0^\delta\M_2$.  If  there exists a Markov policy $\mu$ for $\M_1\otimes \mathcal A_\psi$ such that the accepting states are reached with $\delta$-robust probability $p$, then there exists a control strategy $\Ca_2$ for $\M_2$ such that the accepting states of $\mathcal A_\psi$ are reached with probability $p$ under the evolution of $\Ca_2\times \M_2$. More precisely, denote with $\bar \X:=\X_1\times Q$ the state space of $\M_1\otimes \mathcal A_\psi$, then the mapping $ \mathbf T_\delta^\nu$ becomes
\begin{align}
\notag \mathbf T_\delta^\nu (V)(x_1,q)=\Lim\Big(\int_{\X_1} \!\!\!\!   \left[\mathbf 1_{F}(\trans_x(q,x_1'))+\mathbf 1_{Q\setminus F}(\trans_x(q,x_1')) V(x_1',\trans_x(q,x_1')) \right]\quad \\
\times \mathbb T(dx'_1|x_1,\nu(x_1,q))-\delta\Big)\label{eq:kronmap}\
\end{align}
with $\trans_x(q,x_1'):= \trans(q,\mathsf L(h_1(x_1'))))$. For  $\mathbf T_\delta^\mu (V)(x_1,q)=V(x_1,q)$, the $\delta$-robust reachability probability is defined as
\begin{align}\notag r^\mu(F\times \X_1)= \Lim\Big( \int_{\X_1}[ \mathbf{1}_{F}(\trans_x(q_0,x_1)) +\mathbf 1_{{Q\setminus F}}(\trans_x(q_0,x_1))\hspace{2cm}\\ \times V(x_1,\trans_x(q_0,x_1))]\pi(d x_1) -\delta\Big ) .\end{align}


\subsection{$(\eps,\delta)$-Robust satisfaction of scLTL properties}
\begin{figure}[htp]
  \includegraphics[width=.45\textwidth]{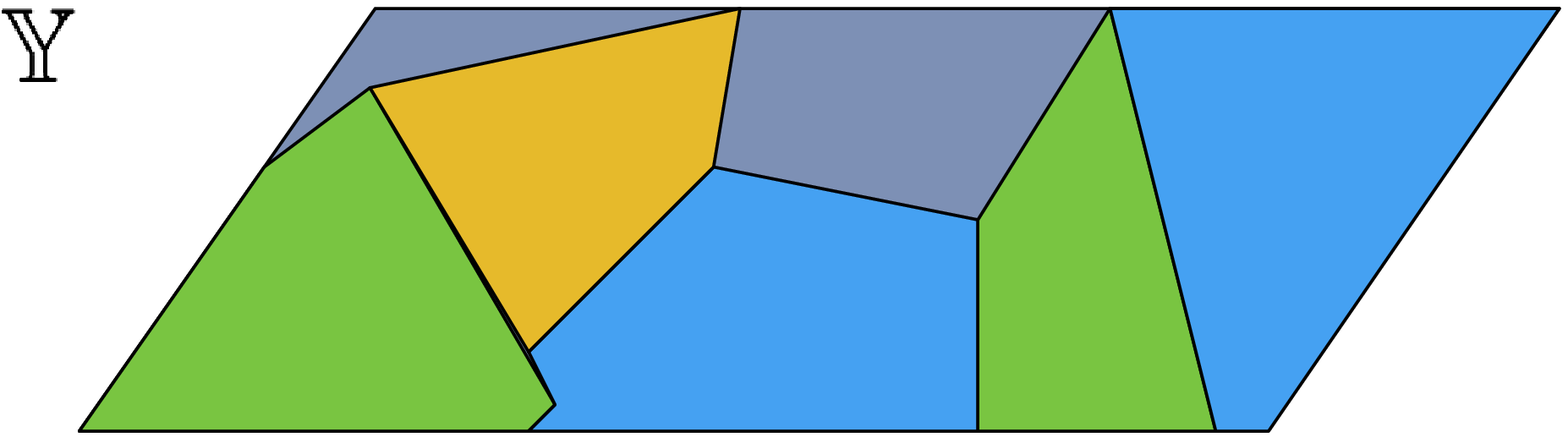}\hfill   \includegraphics[width=.45\textwidth]{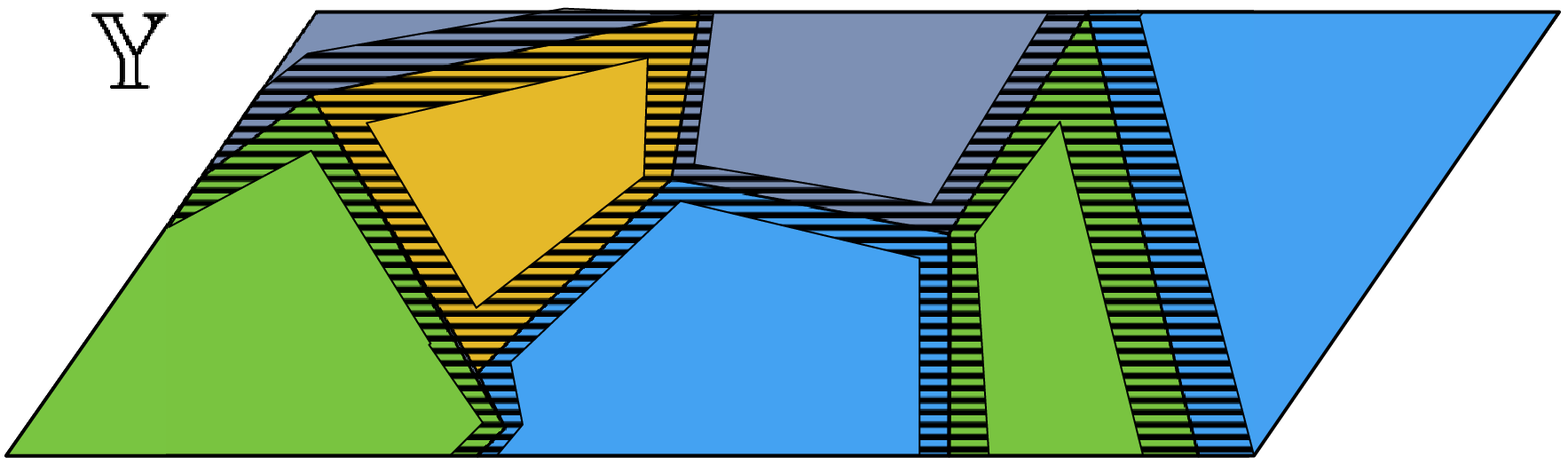}
  \caption{A typical labelling over the output space. On the left a normal labelling is given, on the right the labelling is non-deterministic due to  the output error.}\label{fig:label}
\end{figure}

We now integrate the $\eps$ error in the output space 
into the robust synthesis problem via the  effect it has on the labelling. Given $\Lab:\Y\rightarrow \Sigma$,  we define $\Lab_\eps:\Y\rightarrow 2^{\Sigma}$ as
\begin{equation}
\Lab_\eps(y):=\{q\in\Sigma| \exists (y,y_\eps): d_\Y(y,y_\eps)\leq \eps \mbox{ and } q= \Lab(y_\eps)\}.
\end{equation}
Consider $\M_1\preceq_\eps^\delta\M_2$ with $\rel_\eps$, then for all $(x_1,x_2)\in\rel_\eps$, it holds that
$\Lab(h_2(x_2))\in \Lab_\eps (h_1(x_1))$.  In Figure \ref{fig:label}, an output space together with its labels is depicted. When taking into account an $\eps$ error in the output, the labelling becomes non-deterministic in some regions. This can be observed on the right of Figure \ref{fig:label}.

Instead of integrating this relaxed labelling into the product construction of a given gMDP, we will immediately adapt the $\delta$-robust reachability computations \eqref{eq:kronmap} to deal with this 
non-determinism.
Consider the $(\eps,\delta)$-robust operator
\( \mathbf T_{\eps,\delta}^{\nu}(V)(x_1,q)
\)  defined as 
\begin{align*}\notag \mathbf T_{\eps,\delta}^\nu (V)(x_1,q)=\Lim\Big(\int_{\X_1} \min_{q'\in \bar\trans_x(q,x_1')}[  \mathbf 1_{F}(q')+\mathbf 1_{Q\setminus F}(q') V_{k+1}(x_1',q') ]\times\qquad  \\    \mathbb T(dx'_1|x_1,\nu(x_1,q))-\delta\Big)
\end{align*} 
with $\bar\trans_x(q,x_1):=\{\trans(q,\alpha)\mbox{ with }\alpha \in \Lab_{\epsilon}(h_1(x_1))\}$. For a time-homogeneous Markov policy $\mu$ and $V(x_1,q)$ that satisfy $\mathbf T_{\eps,\delta}^\mu (V)(x_1,q)=V(x_1,q)$, the $\delta$-robust reachability probability is defined as
\begin{equation*}
r^{\eps,\delta}(F\times \X_1)= \Lim\Big( \int_{\X_1}\min_{q'\in \bar\trans_x(q_0,x_1)}[ 1_{F}(q') +\mathbf 1_{{Q\setminus F}}(q')   V(x_1,q')]  \pi(d x_1) -\delta\Big ).
\end{equation*}
Consider a scLTL property $\psi$ and the corresponding $\mc A_\psi$ with goal states $F$. If  $F\times \X_1$
is $\delta$-robust reachable with respect to $r^{\eps,\delta}(F\times \X_1)$, then we can refine $\mu$ to $\Ca_2(\mu,\psi)$ such that  $\psi$ is satisfied by $\Ca_2(\mu,\psi)\times \M_2$ with a probability $p\geq r^{\eps,\delta}(F\times \X_1)$. Of course the apparent non-determinism due to the relaxed labelling will be  resolved in the refined control strategy by selecting the labels of the concrete model.

Building on Corollary \ref{thm:delreach_infty}, we can now also maximise  the  $(\epsilon,\delta)$-robust probability  using $\mathbf T_{\eps,\delta}^{\ast}$, defined as
$ \mathbf T_{\eps,\delta}^{\ast}(V)(x_1,q):= \sup_{\mu}\mathbf T_{\eps,\delta}^{\mu}(V)(x_1,q),
$ 
which yields an optimised robust Markov policy as
\[ \mu^\ast(x_1,q)\in\arg\sup_{\mu}\mathbf T_{\eps,\delta}^{\mu}(V^*)(x_1,q) \mbox{ for }  \mathbf T_{\eps,\delta}^{\ast}(V^*)(x_1,q) = V^*(x_1,q).
\] 
Hence, we have shown that we can leverage probabilistic simulation relations to use approximate models for the   controller synthesis and verification of both probabilistic reachability (cf. Problem \ref{prob:reach}) and scLTL properties (cf. Problem \ref{prob:scLTL}).

\section{Computational implementation and toy example}
\label{sec:case_study}
In this section, we apply our results to formal controller synthesis for stochastic linear dynamical systems.  Existing formal controller synthesis results for this class of models either rely on model order reduction \cite{LSMZ17} or use abstraction techniques as finite-state MDPs \cite{SAID,tmka2013}. Our new results combine these two approaches in one framework that benefits from both of them. More precisely, our abstract model used for synthesis is obtained by discretising the state space of a reduced-order version of the concrete model.

\noindent\textbf{Concrete model.} Consider the following linear dynamical model $\M_2$: 
\begin{align*}
& x_2(t+1) = A_2 x_2(t) + B_2 u_2(t) + B_{w2} w(t),\quad x_2(0) = x_{20}\in\X_2\\
& y_2(t) = C_2 x_2(t),\quad t = 0,1,2,\ldots, 
\end{align*}
where $x_2(\cdot)\in\X_2\subset\mathbb R^n$, $u_2(\cdot)\in \A_2\subset \mathbb R^m$, and $y_2(\cdot)\in\Y\subset\mathbb R^p$. Matrices $A_2$, $B_2$, $B_{w2}$, and $C_2$ have appropriate dimension and $w(\cdot)$ are iid random variables with standard multivariate Gaussian distributions.

\smallskip
\noindent\textbf{Constructing the abstract model.}
Construction of the abstract model relies on partitioning a new space $\X_s\subset \mathbb R^{n_s}$, where $n_s<n$, as $\{\mathbb A_i\subset \X_s,\,\,i=1,2,\ldots,l\}$.  
Over this partition, we select representative points $\{z_i\in \mathbb A_i,\,\,i=1,2,\ldots,l\}$, which we call $\X_1$ and which becomes the state space of the abstract model $\M_1$. 
Introduce the operator $\Pi:\X_s\rightarrow\X_1$ that assigns to any $x_1\in \mathbb A_i$, $i\in\{1,\ldots,l\}$ the representative point of $\mathbb A_i$, $z_i = \Pi(x_1)$.\\* Next we provide a dynamical characterisation of $\M_1$. 
The state evolution of $\M_1$ is written as
\begin{align*}
& x_1(t+1) = \Pi\left(A_1 x_1(t) + B_1u_1(t) + B_{w1}w(t)\right),\quad x_1(0) = x_{10}\in \X_1\\
& y_1(t) = C_1 x_1(t),\quad t=0,1,2,\ldots, 
\end{align*}
with state $x_1(\cdot)\in\X_1$, input $u_1(\cdot)\in\A_1$, and output $y_1(\cdot)\in\Y$, and matrices $A_1,B_1,B_{w1},C_1$ of appropriate dimensions.
Note that the noise term $w(t)$ in $\M_1$ is the same as the one in $\M_2$, thereby allowing to define a lifting $\mathbb W_{\mathbb T}$.

\smallskip
\noindent\textbf{Computing the ($\eps,\delta$)-simulation relation.} Consider the linear interface function $u_2 = R u_1 + Q x_1 + K(x_2-Px_1)$, such that $PA_1 = A_2P + B_2Q$ for some matrix $P$.
Define the relation $(x_1,x_2)\in\rel_\delta^\epsilon$ iff $(x_2-Px_1)^TM(x_2-Px_1)\le \epsilon^2$.
Next we check conditions of Def.~\ref{def:apbsim} under which $\M_1\preceq_\eps^\delta \M_2$.
It is guaranteed that $d_\Y(y_1,y_2)\le \epsilon$ for any $(x_1,x_2)\in\rel_\delta^\epsilon$ (cf. \textbf{APS1} in Def.~\ref{def:apbsim}) if $ C_1 = C_2P$, and $C_2^TC_2\le M$.
Condition \textbf{APS2} in Def.~\ref{def:apbsim} holds if $c_w$ is selected such that $\mathbb P(w^T w\le c_w)\ge 1-\delta$ and the following inequality
	\begin{equation}
	\label{eq:case_study}
	(\bar A\bar x+\bar B u_1 + \bar B_w w + P\beta)^T M(\bar A\bar x+\bar B u_1 + \bar B_w w+P\beta)\le \epsilon^2
	\end{equation}
	is satisfied for any $\bar x,u_1,w,\beta$ such that $w^T w\le c_w$, $u_1^2\le c_u$, $\bar x^TM\bar x\le \epsilon^2$, $|\beta|\le \vec\delta$.
	The matrices in \eqref{eq:case_study} are defined as $\bar A := A_2+B_2K$, $\bar B := B_2R-PB_1$, and $\bar B_{w2} := B_{w2}-PB_{w1}$.
	Vector $\vec\delta$ is the diameter of the partition $\{\mathbb A_i,i=1,\ldots,l\}$, which satisfies $|x_s-x_s'|\le\vec\delta$ component-wise for any $x_s,x_s'\in \mathbb A_i$ and any $i\in\{1,2,\ldots,l\}$. Condition~\eqref{eq:case_study} can be checked 
	using LMIs and the S-procedure \cite{Boyd2004}. 


For 
	Condition~\textbf{APS3} in Def.~\ref{def:apbsim}, we now question when there exists  a deterministic  initial state $x_{10}$ for a given deterministic initial state $x_{20}$ satisfying  $(x_{20}-Px_{10})^TM(x_{20}-Px_{10})\le \epsilon^2$.
	We can choose $x_{10} := \hat P x_{20}$ with $\hat P:=(P^T M P)^{-1} P^T M$ which is the minimum of left-hand side
	, and select it as the representative point of the associated partition set in $\M_1$.
	Alternatively, when the representative point cannot be freely chosen, we select $x_{10}=\Pi(\hat P x_{20})$.\\
	In the former case, there exists an initial state $x_{10}$ if 
	$x_{20}^T 
		M(\mathbb I_n -  P\hat P) x_{20}\le \epsilon^2$, 
	or, dually, $\epsilon$ is lower bounded for a given $x_{20}$.
	In the latter case, if 
		$\|
	M^{\frac{1}{2}}(\mathbb I_n -  P\hat P) x_{20}\|+\|M^{\frac{1}{2}}P\delta\| \le \epsilon$, then there exists an initial state $x_{10}$ satisfying Condition~\textbf{APS3}.
	\subsubsection{Toy example.}
\begin{figure}[htp]
	\includegraphics[width=.45\textwidth]{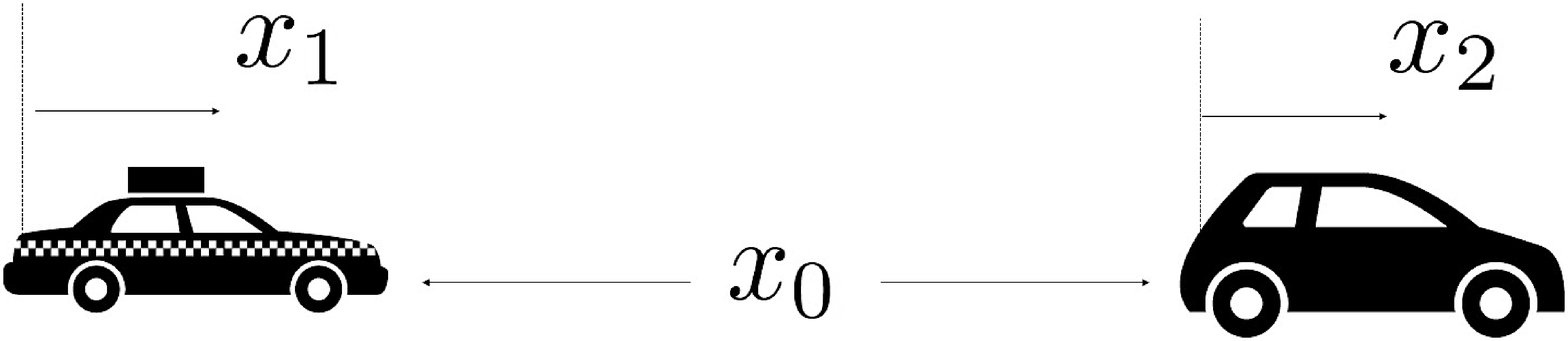}
	\hspace{0.5cm}
	\includegraphics[width=.45\textwidth]{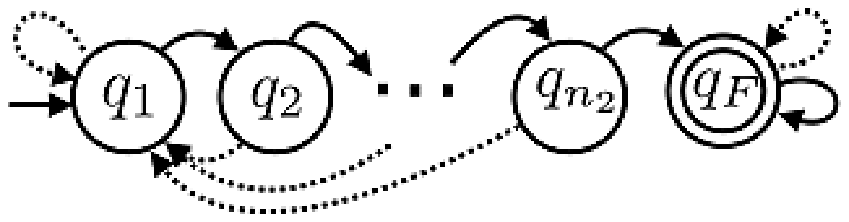}
	\caption{A game of tag:  $\eventually\always^{\le n_2}\{x_a\in K\}$}\label{fig:gametag}\mbox{} \\[-1cm]
\end{figure}

We consider the specification $\psi = \eventually\always^{\le n_2}\{y\in K\}$ which encodes reach and stay over bounded time intervals. The associated DFA is given in Figure \ref{fig:gametag}, together with an illustration of a potential application of this toy example.

Consider the original model $\M_2$, which is a 3 dimensional model with output $y_1(t)=x_a$ and 
\begin{align}
x_a(t+1)&=x_a(t)- a_1(x_b(t)-x_c(t))-a_2u(t)+a_3w(t)\notag
\\
x_b(t+1)&=b x_b(t)+u(t)\notag\\
x_c(t+1)&= c_1x_c(t)+c_2w(t)
\end{align}
with $a_1=.3$, $a_2=0.03$, $a_3=6e\mbox{-}3$, $b=c_1=.8$ and $c_2=.1$.
For the game we consider the case with $n_2=3$. 
\input{Fig01.tex}
As in \cite{haesaert2017verification}, we compute the lower dimensional model via the balance truncation of the original controlled model with a suitable  feedback gain ($K=[-0.7738    0.9369   -0.6829]$). 
%
In Figure \ref{fig:simresult}, we give an example of such a robust temporal logic computation.  On the right side of the figure, 10 simulation runs are given \footnote{initiated at $x_a=2.45$, $x_b=2.5$ and $x_c=1.3$.}, the crosses are the outputs of $\M_1$ whereas the lines are the outputs of $\M_2$.

\section{Conclusions and future work}
In this paper, we have introduced a new robust way of synthesising control strategies and verifying probabilistic temporal logic properties. Beyond this theoretical contribution, future work will focus on the computational aspects of this approach to prepare for application on realistic sized problems.

\bibliographystyle{abbrv}

\appendix
\input{Appendix}
\end{document}

%% file: Commands.tex
\usepackage{graphicx,dblfloatfix,subcaption}
\usepackage{wrapfig}
\usepackage[innercaption]{sidecap}
\usepackage{tikz,epstopdf,pgfplots,pdfpages}
\usepackage{framed}
\usepackage{marginnote}
\usepackage{blkarray}
\usepackage{enumerate}%
\usepackage[author={Sofie},draft]{pdfcomment}
\usepackage{amsmath,amssymb,amsfonts}
\usepackage{mathtools}
\usetikzlibrary{decorations.shapes,shapes.geometric,shadows,arrows,automata,positioning,calendar,mindmap,backgrounds,scopes,chains,er,patterns,pgfplots.groupplots}
 \usetikzlibrary{calc}
\usepackage[nonumberlist]{glossaries}
\newglossary[slg]{symbolslist}{syi}{syg}{List of symbols}

\newcommand{\Lim}{\mathbf L}

\usepackage{algorithm}
\usepackage{algpseudocode}

\newcommand{\word}{\omega}

\newcommand{\mathscr}[1]{\mathcal{#1}}
 \newcommand{\mc}[1]{\mathcal{#1}}

\newtheorem{cor}{Corollary}

    \newtheorem{prob}{Problem}
 \newcommand{\newsymb}[4]{
 \newcommand{#2}{#3}
 \newglossaryentry{symb:#1}{
 name=$#2$,
 description={#4},
 sort=symbol#1, type=symbolslist
 }
 }
 
\newsymb{Hilbert}	{\Hilbert}	{\mathcal{H}_2}
{The Hilbert space}
\newsymb{Reals}		{\Real}		{\mathbb R}
{The set of reals}
\newsymb{Naturals}	{\Nat}		{\mathbf{N}}
{The set of natural numbers}
\newsymb{Complex}	{\Complex}	{\mathbb C}
{The set of complex numbers}

{} 

\newcommand{\po}{\mathbb{P}}     

\newcommand{\norm}[1]{\left\Vert#1\right\Vert}
\newsymb{norm}{\normempty}{\norm{\cdot}}
{The Euclidean norm $\|x\|=\sqrt{x^Tx}.$}

\newcommand{\eps}{\epsilon}

\newsymb{system}{\s}{\mathbf{S}}{The true `(cyber-)physical' system}
\newsymb{model}{\M}{\mathbf{M}}{A mathematical model}
\newsymb{detmodel}{\Md}{\bar{\mathbf{M}}}{The minimal realisation of $\M$ when the noise input is neglected.}

\newsymb{statespace}{\X}{\mathbb{X}}{The state space, i.e. the set of states}     
	\newsymb{state}{\xv}{\mathbf{x}}{An element of the state space $\X$}
\newsymb{actionspace}{\A}{\mathbb{U}}{The set of controllable inputs}
	\newsymb{actions}{\acv}{u}{An element of the set of controllable input $\U$}
\newsymb{outputspace}{\Y}{\mathbb{Y}}{The set of output values}     
    \newsymb{output}{\yv}{y}{An element of the set of outputs}
\newsymb{performanceoutput}{\Z}{\mathbb{Z}}{The set of performance outputs}     

\newsymb{observer}{\Obs}{\mathbf{O}}{An observer}

\newsymb{TS}{\TS}{\Sigma}{A transition system}

	\newsymb{XTS}{\XTS}{\mathscr{X}}{The set of states of a transition system}
	\newsymb{xTS}{\xTS}{x}{The set of states of a transition system}
	\newsymb{UTS}{\UTS}{\mathscr{A}}{The set of inputs to a transition system}
	\newsymb{uTS}{\uTS}{u}{The set of inputs to a transition system}
	\newsymb{ZTS}{\ZTS}{\mathscr{Z}}{The set of possible infinite observations}
	 \newsymb{HTS}{\HTS}{\mathscr{H}}{The labelling map}

\newsymb{initialstates}{\init}{\mathbb{I}}{The set of initial states}

\newsymb{Markov}{\Markov}{\mathcal{M}}{A Markov process}
\newsymb{Relation}{\rel}{\mathcal{R}}{A relation}
\newsymb{EquivRelation}{\erel}{\mathcal{R}}{An equivalence  relation}
\newcommand{\trans}{\mathsf{t}}

\newsymb{similated}{\simu}{\preceq_{\mathcal{S}}}{Simulated by}
\newsymb{bisimilar}{\bsimu}{\sim_{\mathcal{B}}}{Bisimulation equivalence}
\newsymb{simulationequiv}{\esimu}{\simeq_{\mathcal{S}}}{Simulation equivalence}
\newsymb{altsimilated_app}{\asimu}{\preceq^\eps_{\mathcal{AS}}}{$\eps$-approximate alternating simulation}
\newsymb{similated_app}{\simuapp}{\preceq^\eps_{\mathcal{S}}}{$\eps$-approximate simulated by}

\newsymb{until}{\un}{\mbox{\textbf{U}}}{until operator}
\newcommand{\always}{\Box}
\newcommand{\eventually}{\Diamond}
\newcommand{\until}{\mathbin{\sf U}}

\newcommand{\nex}{\mathord{\bigcirc}}

\newcommand{\alphabeth}{\Sigma}

\newcommand{\Lab}{\mathsf{L}}   

\newcommand{\AP}{AP}            

\newcommand{\InF}{\mathcal{U}_{v}}

\newcommand{\Ca}{{\mathbf{C}}}

\newcommand{\cut}[1]{}

\newcommand{\red}[1]{{\color{pink!60!red}#1 \color{black}}}

\newcommand{\removed}[1]{ \marginpar[\hfill{ \tiny [\red{deleted text}]}]{\tiny [\red{deleted text}]}}

\colorlet{shadecolor}{black!10}

\usepackage{mathtools}

\makeatletter
\newcommand\tabfill[1]{%
\dimen@\linewidth%
\advance\dimen@\@totalleftmargin%
\advance\dimen@-\dimen\@curtab%
\parbox[t]\dimen@{#1\ifhmode\strut\fi}%
}
\newcommand{\Wt}{\mathbb{W}_{\mathbb{T}}}
\newcommand{\Wtp}{\bar{\mathbb{W}}_{\mathbb{T}}}
\newcommand{\pcm}[2]{\po_{\scalebox{0.5}[.5]{$\!\!#1\!\!\times\!\!#2$}}}

%% file: ProblemsStatement.tex
\section{Problem statement: temporal logic control}
 \subsection{Preliminaries and notations}

%

In this work, we only consider  Borel measurable spaces, i.e., $(\X,\mathcal{B}(\X))$, 
and we restrict our attention to Polish spaces \cite{bogachev2007measure}. 
Together with the measurable space $(\X,\mathcal{B}(\X))$,  a probability measure $\po$ defines the probability space, denoted by $(\X,\mathcal{B}(\X),\po)$ and has realisations  $x\sim \po$.   
Let us further denote the set of all probability measures for a given measurable space $(\X,\mathcal{B}(\X))$ as $\mathcal P (\X,\mathcal{B}(\X))$.

For the sets $A$ and $B$ a relation $\rel\subset A\times B$ is a subset of the Cartesian product $A\times B$. The relation $\rel$ relates $x\in A$ with $y\in B$ if $(x,y)\in\rel$, which is equivalently written as $x\rel y$.
For a given set $\Y$ a metric or distance function $\mathbf d_\Y$ is a function $\mathbf{d}_\Y: \Y\times \Y\rightarrow \mathbb R_{\ge 0}$ 
satisfying the following conditions: 
$\forall y_1,y_2,y_3\in\Y$:
$\mathbf d_\Y(y_1,y_2)=0$ iff $y_1=y_2$; 
$\mathbf d_\Y(y_1,y_2)=\mathbf d_{\Y}(y_2,y_1)$;  and
$\mathbf d_\Y(y_1,y_3)\leq \mathbf d_\Y(y_1,y_2) +\mathbf d_\Y(y_2,y_3)$.

\subsection{General Markov decision processes and control strategies}
General Markov decision processes are related to control Markov processes \cite{Abate2011} and Markov decision processes \cite{Bible,mt1993,hll1996}, and formalised as follows.
\begin{definition}[general Markov decision process (gMDP)]
\label{def:MDP}
  A discrete-time gMDP is a tuple $\M\!=\!(\X,\!\pi,\!\mathbb T,\!\A,\!h,\! \Y)$ with\\[-2em]
  \begin{itemize}
    \item $\X$,  an (uncountable) Polish state space with states $x\in\X$ as its elements;
       \item $\A$,  the set of controls which is a Polish  space;
    \item $\pi$, the initial probability measure $\pi:\mathcal{B}(\X)\rightarrow [0,1]$;
    \item $\mathbb T:\X\times\A\times\mathcal B(\X)\rightarrow[0,1]$, a conditional stochastic kernel assigning to each state $x\in \X$ and control $u\in \A$  probability measure $\mathbb T(\cdot\mid x,u)$ over $(\X,\mathcal B(\X))$;
    \item $\Y$, the output space decorated with metric $\mathbf d_\Y$; and
    \item $h:\X\rightarrow\Y$, a measurable output map.
  \end{itemize}
\end{definition}
For any set $A\in \mathcal{B}(\X)$, $\po_{x,u}(x(t+1)\in A)=\int_A \mathbb T(d x'\mid x(t)=x,u)$, where $\po_{x,u}$ denotes the conditional probability $\po(\cdot\mid x,u)$.
At every state, the state transition depends non-deterministically on the choice of $u\in \A$.
When chosen according to
a probability measure  $\mu_u:\mathcal{B}(\A)\rightarrow [0,1]$, we refer to the stochastic control input as $\mu_u$ and denote
the transition kernel as $\mathbb T(\cdot| x, \mu_u)=\int_\A \mathbb T(\cdot| x, u) \mu_u(du)\in \mathcal P(\X,\mathcal B(\X))$.
Given a string of inputs
$\{u(t)\}_{t\le N}:=u(0), u(1), \ldots, u(N)$,
over a finite time horizon $[0,N]$,
and an initial condition  $x_0$ (sampled from $\pi$),
the state at the $(t+1)$-st time instant, $x(t+1)$,
is obtained as a realisation of the controlled Borel-measurable stochastic kernel $\mathbb{T}\left(\cdot\mid x(t), u(t) \right)$ --
these semantics induce paths (or executions) of the gMDP.
Further, output traces of gMDP are obtained by applying the output map $h(\cdot)$ to the paths of gMDP, namely
$\{y(t)\}_{t\le N}:= y(0), y(1), \ldots, y(N)$ with $y(t) = h\big(x(t)\big)$ for all $t\in[0,N]$.
Denote the
class of all gMDP with the same metric output space $\Y$ as $\mathcal{M}_\Y$. 


A policy is a selection of control inputs based on the past history of states and controls.
When the selected controls are only dependent on the current states,
the policy is referred to as Markov.
\begin{definition}[Markov policy]
\label{def:markovpolicy}
For a gMDP $\M=(\X,\pi,\mathbb T,\A, h, \Y)$, a Markov policy $\mu$ is a sequence $\mu=(\mu_0,\mu_1,\mu_2,\ldots)$ of universally measurable maps $\mu_t=\X\rightarrow \mathcal P(\A,\mathcal B(\A))$, $t\in\mathbb N:=\{0,1,2,\ldots\}$, from the state space $\X$ to the set of controls.
\end{definition}
We allow controls to be selected via universally measurable maps \cite{Bible} from the state to the space of stochastic control inputs,
so that properties such as safety can be maximised \cite{Abate1}.
%
%
%
We introduce the notion of a control strategy, and define it as a broader, memory-dependent version of the Markov policy above.
This strategy is formulated again as a gMDP that takes as its input the state of the to-be-controlled gMDP.
\begin{definition}[Control strategy]
\label{def:CS}
A control strategy $\Ca=(\X_\Ca,x_{\Ca0},\X,\mathbb T_\Ca,h_\Ca)$ for a gMDP $\M=(\X,\pi,\mathbb T,\A, h, \Y)$ is a gMDP  with
state space $\X_\Ca$;
initial state $x_{\Ca 0}$;  input space $\X$;
universally measurable kernels $\mathbb T_{\Ca}:\X_\Ca\times\X\times\mathcal B(\X_\Ca)\rightarrow[0,1]$; and with universally measurable output map $h_\Ca:\X_\Ca\rightarrow \mathcal P(\A,\mathcal B(\A))$.
 %
\end{definition} 
Note that the stochastic transitions for the control strategy and the gMDP are selected in an alternating fashion.
 The output map of the strategy is indexed based on the time instant at which the resulting policy will be applied to the gMDP. This is further elucidated in Algorithm~\ref{alg:CM}.
 \begin{algorithm}\caption{Execution semantics of the controlled model $\mathbf C\times \M$. 
 }
\label{alg:CM}
 \begin{minipage}[c]{.6\textwidth}
   \begin{algorithmic}[1]
           \Procedure{Execution}{$\Ca$, $\M$}  
\State set $t:=0$ and $x_\Ca (0):=x_{\Ca 0}$ 
\State {draw $x(0) \sim \pi$ }
\For{$t\leq N$}
\State{draw  $x_\Ca(t+1) \sim \mathbb T_\Ca(\,\cdot\,{\mid} x_\Ca(t),x(t) )$ }
\State{draw $u(t)$ from $\mu_t:=h_\Ca(x_\Ca(t+1))$}
\State{draw $x(t+1) \sim \mathbb T(\,\cdot\,{\mid}x(t),u(t) ))$ }
\State{set $t:=t+1$}
\EndFor
\State 
\Return  $\{x(t)\}_{t\le N}$
\EndProcedure
\end{algorithmic}
\end{minipage}
\begin{minipage}[c][4.5cm]{.27\textwidth}
  \mbox{ }\\[1em]
 \resizebox{3.75cm}{!}{ \begin{tikzpicture}[ node distance=1.5cm]
    \tikzset{cstate/.style={diamond,draw, fill=yellow!80!gray!40, inner sep=.1cm}}
    \tikzset{state/.style={circle, draw, fill=blue!40, inner sep=.1cm}}
    \node[cstate] (cx0) {};
    \node[left of=cx0,yshift=2em,node distance=1.5cm] (l1) {};
    \node[state,  right of = cx0, yshift=-0cm] (x0){};  

       \node[ above of =x0,yshift=-.75cm] (x0pi) {$\pi$};
    \node[right of=x0pi, yshift=.2em, node distance=.75cm] (M_text) {$\M$};%
	\node[left of=cx0, yshift=.2em, node distance=.75cm] (C_text) {$\Ca$};%

    \node[cstate, below of= cx0] (cx1) {};
    \node[state,  below of = x0] (x1) {};

    \node[cstate, below of= cx1] (cx2){};
    \node[state,  below of = x1] (x2){};
  \node[cstate, below of= cx2] (cx3){};
  \node[  below of = x2, node distance=.75cm] (x3){$\vdots$};
    \node[left of=cx3,yshift=0em,node distance=1.5cm] (l2) {};

    \path[->,draw] (x0pi)->(x0);
    \path[->,draw] (x0)edge node [right] {\small $x(0)$} (x1);
    \path[->,draw] (x1)edge node [right] {\small$x(1)$} (x2);

    \path[->,draw] (cx0) edge node [left] {\small$x_\Ca(0)$} (cx1);
    \path[->,draw] (cx1)edge node [left] {\small$x_\Ca(1)$} (cx2);
    \path[->,draw] (cx2)edge node [left] {\small$x_\Ca(2)$} (cx3);
    \path[->,draw, dashed] (x0)edge node [state, fill=white ,draw=white, inner sep=0] {\small$x(0)$} (cx1);
    \path[->,draw, dashed ] (x1)edge node [state, fill=white ,draw=white, inner sep=0] {\small $x(1)$ }(cx2);
    \path[->,draw, dashed ] (x2)edge node [state, fill=white ,draw=white, inner sep=0] {\small $x(2)$ }(cx3);
    \path[->,draw, dashed ] (cx1) edge node[state, fill=white ,draw=white, inner sep=0]{\small $u(0)$} (x1);
    \path[->,draw, dashed ] (cx2) edge node[state, fill=white ,draw=white, inner sep=0]{\small $u(1)$} (x2);
    \path[draw, ->] (x2)->(x3);
    \path[draw,dashed] (l1)->(l2);
    \node[right of =x0,node distance=1cm](y0){$y(0)$};
        \node[right of =x1, node distance=1cm](y1){$y(1)$};
        \node[right of =x2,node distance=1cm](y2){$y(2)$};
        \path[draw, ->] (x0)->(y0);
        \path[draw, ->] (x1)->(y1);
        \path[draw, ->] (x2)->(y2);

  \end{tikzpicture}}
\end{minipage} \mbox{ }\\[1em]
 \end{algorithm}
The execution \mbox{$\{(x(t), x_\Ca(t)), t\in[0,N]\}$} of a gMDP $\M$ controlled with strategy $\Ca$ (denoted by $\Ca\times \M$)
is defined on the canonical sample space $\Omega:= (\X\times\X_\Ca)^{N+1}$ endowed with its product topology $\mathcal B (\Omega)$ and with  a 
unique probability measure $\pcm{\Ca}{\M}$.
A Markov policy is as a special case of control strategy 
which does not have an internal state that can be used to remember relevant past events. 
\begin{remark}Any Markov policy $\mu$ as defined in Def.~\ref{def:markovpolicy}  can be written as a control strategy $\Ca_\mu:=(\X_\mu,x_{\mu 0},\X,\mathbb T_\mu,h_\mu)$  with 
	$\X_\mu:= Q\times \X$, for which $Q$ is the set of time indices $Q:=\{-1,0,1,2,3,\ldots\}$, and with
	$x_{\mu 0}:=(-1,x_0)$ the initial state for some $x_0\in \X$.
	The probability measure on the next control state $x_\mu'=(q',\tilde x')\in \X_\mu$ given $x(t)$ and given the current state $x_\mu=(q,\tilde x)$ is defined with a stochastic kernel as
 \begin{align*}\textstyle
  \mathbb T_\mu(A|(q,\tilde x),x(t)) :=\left\{\begin{array}{l}
  1 \mbox{ if } (q
  +1,x(t))\in A ,\\
  0 \mbox{  else } .
  \end{array}\right.
  \end{align*}
  The policies can then be embedded into the control strategy via the output map  $h_\mu((q,\tilde x)):=\mu_q(\tilde x)$.\footnote{ Note that for this construction, the separability of the Polish space $\X$ is important as otherwise $  \mathbb T_\mu$ would not be measurable in general.}
\end{remark}
 

\subsection{Probabilistic path properties of controlled gMDPs }
   
Consider a measurable target set $K\subset \mathbb Y$.
We say that an output trace $\{y(t)\}_{t\leq N}$  reaches a target set $K$, if there exists a time $t\in [0,N]$ such that $y(t)\in K$.
This bounded reaching of $K$ is denoted by $ \lozenge^{\le N}\{y\in K\}$ or briefly $\lozenge^{\le N}K$.  For $N\rightarrow \infty$, we denote  the reachability property as $ \lozenge K$, i.e., eventually $K$.
For a given gMDP $\M$ with control strategy $\Ca$,
a verification task consists of quantifying  
the probability that an output trace of $\Ca\times \M$ reaches $K$ within the time horizon $[0,N]$ , i.e., 
\( \pcm{\Ca}{\M}(\lozenge^{\le N} K)\),
or that the target set $K$ is eventually reached, i.e., 
\(
  \pcm{\Ca}{\M}(\lozenge K)
\), and verifying that it is within a given threshold.




More complex properties can be described using temporal logic.
Consider a set of atomic propositions $\AP$, the alphabet $\alphabeth := 2^{\AP}$, and
infinite words that are string composed of elements from $\alphabeth$, $\word=\word(0),\word(1),\word(2),\ldots\in\alphabeth^{\mathbb{N}}$.
Of interest are atomic propositions that are connected to the gMDP via a measurable labelling function $\Lab:\mathbb Y\rightarrow \alphabeth$ from the output space to the alphabet $\alphabeth$. 
Via its trivial extension,  output traces  $\{y(t)\}_{t\geq 0}\in \Y^{\mathbb N} $ can be mapped to the set of infinite words $\alphabeth^{\mathbb N}$, as 
$\word=\Lab(\{y(t)\}_{t\geq0}) := \{\Lab(y(t))\}_{t\geq0}$.
%
Consider linear-time temporal logic properties with syntax
\begin{equation}
	\psi ::=  \operatorname{true} \,|\, p \,|\, \neg \psi \,|\,\psi_1 \wedge \psi_2 \,|\, \nex \psi \,|\, \psi_1\until \psi_2.
\end{equation}
Let $\word_t=\word(t),\word(t+1),\word(t+2),\ldots  $ be a postfix of the word $\word$, then  
the satisfaction relation between $\word$ and a property $\psi$, expressed via LTL,  is denoted by \(\word\vDash\psi\)  
(or equivalently \(\word_0\vDash\psi\)). 
The semantics of the satisfaction relation are defined recursively over $\word_t$ and
 the syntax of the LTL formula $\psi$.
 An atomic proposition $ p\in \AP$ is satisfied by $\word_t$, i.e.,  $\word_t\vDash p$, iff   $p \in\word(t)$.  Furthermore, 
$\word_t\vDash \neg \psi$  if $\word_t\nvDash\psi$ and 
 we say that  $\word_t\vDash \psi_1\wedge\psi_2$ 
 if $ \word_t\vDash \psi_1$ and if $\word_t\vDash \psi_2$.
The next operator $\word_t\vDash\nex\psi $ holds if the property holds at the next time instance   $ \word_{t+1}\vDash \psi$.  
The temporal until operator $\word_t\vDash \psi_1\until\psi_2$  holds if $ \exists i \in \mathbb{N}:$ $\word_{t+i} \vDash \psi_2$, and 
$\forall j \in\mathbb{N}: 0\leq j<i, \word_{t+j}\vDash \psi_1$.
Based on these semantics, operators such as disjunction ($\vee$) can also be defined through the negation and conjunction:
$ \word_t\vDash \psi_1\vee\psi_2\ \Leftrightarrow  \  \word_t\vDash \psi_1 \mbox{ or } \word_t\vDash \psi_2$.

We are interested in a fragment of LTL property known as syntactically co-safe temporal logic (scLTL) \cite{KupfermanVardi2001}. Even though scLTL formulas are interpreted over infinite words, their satisfaction is guaranteed in finite time. This fragment is defined 
as follows.
\begin{definition}
\label{def:scLTL}
	A scLTL over a set of atomic propositions $\AP$ has syntax 
	\begin{equation}
	\label{eq:scLTL}
	\psi ::=  \operatorname{true} \,|\, p \,|\, \neg p \,|\,\psi_1 \wedge \psi_2\,|\,\psi_1 \vee \psi_2 \,|\, \nex \psi \,|\, \psi_1\until \psi_2\,|\, \lozenge \psi_2
	\end{equation}  
	with $p\in \AP$. 
\end{definition}
In the remainder, we will mainly consider scLTL properties since their verification can be computed via a reachability property over a finite state automaton \cite{KupfermanVardi2001}. 
With respect to a scLTL property $\psi$, we say that a gMDP $\M$ satisfies $\psi$ for a given control strategy $\Ca$ with probability at least $p$ iff 
$	\mathbb P_{\Ca\times\M} (\Lab(\{y(t)\}_{t\geq 0})\models\psi )\geq p$. Apart from this verification task, we are mostly interested in synthesising control strategies such that $\Ca\times\M$ satisfies the inequality.

\subsection{Problem statement}
In this paper, we tackle the control synthesis  for the  (bounded) probabilistic reachability problem and for the temporal logic control problem, defined next.
\begin{prob}[(Bounded) probabilistic reachability] \label{prob:reach}
Given a gMDP $\M$ and a set $K\subset\Y$,
compute a control strategy $\Ca$ that maximises the bounded reachability probability $\pcm{\Ca}{\M}(\eventually^{\leq N} K )$ or the reachability probability $\pcm{\Ca}{\M}(\eventually K )$.
\end{prob}

\begin{prob}[Temporal logic control]\label{prob:scLTL}
	Given a gMDP $\M$, a scLTL property $\psi$ and a labelling function $\Lab$, compute a control strategy  $\Ca$ that maximises the probability of controlled Markov process $\Ca\times \M$ satisfying $\psi$, i.e.,
	\begin{equation}
	\label{eq:prob}
	\max_{\Ca}\,\,\pcm{\Ca}{\M}(\Lab(\{y(t)\}_{t\geq 0})\models\psi ).
	\end{equation}
\end{prob}
Computation  of \eqref{eq:prob} for a given controlled Markov process is generally impossible. In the next sections, we give a robust computations based on \mbox{$(\eps,\delta)$-probabilistic} simulation relations between a given model and its approximation.
%

%% file: DFA.tex
We extend the results on robust probabilistic reachability to the scLTL properties in Def.~\ref{def:scLTL}.  
 For this purpose,  we introduce a model known as Deterministic Finite-state Automaton (DFA).
\begin{definition}[DFA]
  A DFA is a tuple $\mathcal A = \left(Q,q_0,\Sigma,F,\trans\right)$, where
  $Q$ is a finite set of locations,
  $q_0\in Q$ is the initial location,
  $\Sigma$ is a finite set,
  $F\subseteq Q$ is a set of accept locations, and
  $\trans: Q\times\Sigma\rightarrow Q$ is a transition function.
\end{definition}

 A finite word composed of letters of the alphabet defined by $\Sigma$, i.e., $\word = (\word(0),\ldots,\word(n))\in \Sigma^n$, 
is accepted by a DFA $\mathcal A$ if there exists a finite run $q =(q(0),\ldots,q(n + 1))\in Q^{n+2}$ such that $q(0) = q_0$,
$q(i + 1) =\trans(q(i),\word(i))$ for all $0\le i\le n$ and $q(n + 1)\in F$.
Similarly, we say that an infinite word $w\in\mathfrak S$ is accepted by a DFA $\mathcal A$ if there exists a finite prefix of $\word$ accepted by $\mathcal A$ as a finite word. 
More precisely, an infinite word $\word\in\Sigma^{\mathbb{N}_0}$ is accepted by $\mathcal A$ if and only if there exists an infinite run $q\in Q^{\mathbb N_0}$
such that $q(0) = q_0$, $q(i + 1) = \trans(q(i),\word(i))$ for all $i\in\mathbb N_0$ and
there exists $j\in\mathbb N_0$ such that $q(j)\in F$. 
The accepted language of $\mathcal A$, denoted $\mathcal L(\mathcal A)$, is
the set of all words accepted by $\mathcal A$. For every scLTL property $\psi$, c.f. Definition \ref{def:scLTL}, there exists a DFA $\mathcal A_{\psi}$ such that 
\begin{equation}
\word\vDash\mathcal \psi \,\,\Leftrightarrow\,\,  \word\in \mathcal L(\mathcal A_\psi).
\end{equation} 

As a result, the satisfaction of the property $\psi$  now becomes equivalent to the reaching of the accept locations in the DFA. 
We use the DFA $\mathcal A_{\psi}$ to specify properties of the gMDP $\M=(\X, \pi, \mathbb{T}, \mathbb U, h, \mathbb Y)$ as follows.
Remember that $\mathsf L : \Y\rightarrow \Sigma$ is a given measurable function. To each output $y\in \Y$ it
assigns the letter $\mathsf L(y)\in\Sigma$. Given a control strategy $\Ca$, we can define the probability that a path of $\M$ satisfies a scLTL property $\psi$, i.e. $\mathbb P_{\Ca\times \M}
(\omega \in\mathcal L(\mathcal A_\psi))$. 


We can
reduce the computation of $\mathbb P_{\Ca\times \M}
(\omega \in\mathcal L(\mathcal A_\psi))$  over the traces $\omega$ of $\M$ to the reachability problem
 over another gMDP 
 $\M\otimes\mathcal A_\psi$, which we refer to as a product of the gMDP $\M$ and the automaton $\mathcal A_\psi$. This was originally derived in \cite{tmka2013} for MDPs. We give a similar definition of the product construction as.
\begin{definition}[Product between automaton and gMDP]
\label{def:product}
Given a gMDP $\M=(\X,\pi,\mathbb T,\A,h,\Y)$,
a finite alphabet $\Sigma$,
a labelling function $\mathsf L:\Y\rightarrow\Sigma$
and a DFA $\mathcal A_\psi = (Q,q_0,\Sigma,F,\trans)$,
we define the product between $\M$ and $\mathcal A_\psi$ to be another gMDP denoted as
$\M\otimes\mathcal A_\psi = (\bar{\X},\bar\pi,\bar{\mathbb T},\A,\bar h,\Y)$.
Here $\bar{\X} = \X\times Q$, $\bar h(x,q) = h(x)$ for any $(x,q)\in\bar{\X}$, and
\begin{equation*}
  \bar{\mathbb T}(A\times\{q'\}|x,q,u) = \int_{\tilde x\in A}\mathbf{1}(q' = \trans(q,\mathsf L(h(\tilde x))))\cdot\mathbb T(d\tilde  x|x,u).
\end{equation*}
initialised with  $\bar\pi(dx,q) = \pi(dx) \trans(q_0,\mathsf L(h(x))))$. 
\end{definition}

The quantity $\mathbb P_{\Ca\times \M}
(\omega \in\mathcal L(\mathcal A_\psi))$ 
can be related to the reachability probability over the gMDP $\M\otimes\mathcal A$ with a goal state
$G := \X\times F$, as it was shown to be the case for MDPs in \cite{tmka2013}.
\begin{lemma}\label{lem:reachtoprob}
Given gMDP $\M$, alphabet $\Sigma$,
labelling function $\mathsf L$,
and scLTL specification $\psi$ modelled with DFA $\mathcal A_\psi$,
it holds that
\begin{equation*}
	\mathbb P_{\Ca(\mu,\psi)\times \M}
(\omega \in\mathcal L(\mathcal A_\psi)) = \mathbb P_{\mu\times (\mathcal{A}_\psi \otimes \M)}
(\eventually G))
\end{equation*}
%
For any Mrkov policy $\mu$ on the product space of $\mathcal{A}_\psi \otimes \M$
and any control strategy $\Ca(\mu,\psi)$ on $\M$
with properly defined mappings  
between $\mu$ and the control strategy.
\end{lemma}

\subsection{$\delta$-Robust satisfaction of scLTL properties
}

%% file: Fig01.tex
%
%
\begin{figure}
\centering
\definecolor{mycolor2}{rgb}{0.85000,0.32500,0.09800}%
\definecolor{mycolor3}{rgb}{0.92900,0.69400,0.12500}%
\definecolor{mycolor4}{rgb}{0.49400,0.18400,0.55600}%
\definecolor{mycolor5}{rgb}{0.46600,0.67400,0.18800}%
\definecolor{mycolor6}{rgb}{0.30100,0.74500,0.93300}%
\definecolor{mycolor7}{rgb}{0.63500,0.07800,0.18400}%

\definecolor{mycolor1}{rgb}{0.00000,0.44700,0.74100}%
\begin{tikzpicture}
\begin{axis}[%
width=.4\textwidth,
height=2cm, 
scale only axis,
xmin=-4,
xmax=4,
ymin=0,
ymax=1,
axis background/.style={fill=white},ylabel={$r
^{\eps,\delta}(F,\X)$},xlabel={$x_1$},ylabel style={at={(-0.1,0.5)}, anchor=south}
]
\addplot [color=mycolor1, forget plot]
  table[row sep=crcr]{%
-9.95	0.677144666152418\\
-9.85	0.678305421981977\\
-9.75	0.679478388359084\\
-9.65	0.680661257440357\\
-9.55	0.681851665446273\\
-9.45	0.683047342009643\\
-9.35	0.684246268721207\\
-9.25	0.685446837811119\\
-9.15	0.686648000684952\\
-9.05	0.687849395269387\\
-8.95	0.689051440846107\\
-8.85	0.69025538959911\\
-8.75	0.691463325743889\\
-8.65	0.692678105755258\\
-8.55	0.693903236682071\\
-8.45	0.695142693816228\\
-8.35	0.696400684004567\\
-8.25	0.697681366311799\\
-8.15	0.698988547058625\\
-8.05	0.700325370980316\\
-7.95	0.701694033921699\\
-7.85	0.703095544668799\\
-7.75	0.704529563717926\\
-7.65	0.705994344421724\\
-7.55	0.7074867966637\\
-7.45	0.709002685070524\\
-7.35	0.710536962956783\\
-7.25	0.712084229809002\\
-7.15	0.713639284799467\\
-7.05	0.715197733186703\\
-6.95	0.716756588408367\\
-6.85	0.718314801923146\\
-6.75	0.719873647593133\\
-6.65	0.721436890361526\\
-6.55	0.723010682168515\\
-6.45	0.724603151887945\\
-6.35	0.72622368982775\\
-6.25	0.727881968954847\\
-6.15	0.729586790277606\\
-6.05	0.731344882751639\\
-5.95	0.73315982184389\\
-5.85	0.735031248144028\\
-5.75	0.736954561099739\\
-5.65	0.738921227876835\\
-5.55	0.740919780625773\\
-5.45	0.742937477254825\\
-5.35	0.744962477601931\\
-5.25	0.746986253840666\\
-5.15	0.749005834201104\\
-5.05	0.751025402977378\\
-4.95	0.753056782942857\\
-4.85	0.755118439597248\\
-4.75	0.757232881390596\\
-4.65	0.759422668655986\\
-4.55	0.761705636131197\\
-4.45	0.764090296204619\\
-4.35	0.7665726186821\\
-4.25	0.769135376462761\\
-4.15	0.771750922236965\\
-4.05	0.774387574913598\\
-3.95	0.777018779294415\\
-3.85	0.779633018413099\\
-3.75	0.782241435843712\\
-3.65	0.784879766872202\\
-3.55	0.787602016799537\\
-3.45	0.790465618351263\\
-3.35	0.793511206697827\\
-3.25	0.796743630861334\\
-3.15	0.800122856048606\\
-3.05	0.803572496094751\\
-2.95	0.807008821923898\\
-2.85	0.810384048310837\\
-2.75	0.813726308813308\\
-2.65	0.817150644684607\\
-2.55	0.820819922048811\\
-2.45	0.82485852142605\\
-2.35	0.829258169462602\\
-2.25	0.833841884672989\\
-2.15	0.838344117021138\\
-2.05	0.842609843784117\\
-1.95	0.846808088803896\\
-1.85	0.851438258192076\\
-1.75	0.856955591684379\\
-1.65	0.863221580619784\\
-1.55	0.869373639877674\\
-1.45	0.874548455559123\\
-1.35	0.879304163631906\\
-1.25	0.886134685959379\\
-1.15	0.895871226231319\\
-1.05	0.904509041660057\\
-0.95	0.908727712011937\\
-0.850000000000001	0.909833408427297\\
-0.75	0.93997618858648\\
-0.649999999999999	0.939999054673196\\
-0.55	0.93999997723423\\
-0.45	0.939999999506008\\
-0.350000000000001	0.939999999984391\\
-0.25	0.939999999999453\\
-0.149999999999999	0.939999999999982\\
-0.0499999999999998	0.939999999999988\\
0.0499999999999998	0.939999999999988\\
0.149999999999999	0.939999999999982\\
0.25	0.939999999999453\\
0.350000000000001	0.939999999984391\\
0.45	0.939999999506008\\
0.55	0.93999997723423\\
0.649999999999999	0.939999054673196\\
0.75	0.93997618858648\\
0.850000000000001	0.909833408427297\\
0.95	0.908727712011937\\
1.05	0.904509041660058\\
1.15	0.895871226231319\\
1.25	0.886134685959379\\
1.35	0.879304163631906\\
1.45	0.874548455559123\\
1.55	0.869373639877674\\
1.65	0.863221580619784\\
1.75	0.856955591684379\\
1.85	0.851438258192076\\
1.95	0.846808088803896\\
2.05	0.842609843784117\\
2.15	0.838344117021138\\
2.25	0.833841884672989\\
2.35	0.829258169462602\\
2.45	0.82485852142605\\
2.55	0.820819922048811\\
2.65	0.817150644684607\\
2.75	0.813726308813308\\
2.85	0.810384048310838\\
2.95	0.807008821923898\\
3.05	0.803572496094752\\
3.15	0.800122856048606\\
3.25	0.796743630861334\\
3.35	0.793511206697826\\
3.45	0.790465618351263\\
3.55	0.787602016799537\\
3.65	0.784879766872202\\
3.75	0.782241435843712\\
3.85	0.779633018413099\\
3.95	0.777018779294415\\
4.05	0.774387574913598\\
4.15	0.771750922236965\\
4.25	0.769135376462761\\
4.35	0.7665726186821\\
4.45	0.764090296204619\\
4.55	0.761705636131197\\
4.65	0.759422668655986\\
4.75	0.757232881390596\\
4.85	0.755118439597248\\
4.95	0.753056782942857\\
5.05	0.751025402977378\\
5.15	0.749005834201104\\
5.25	0.746986253840667\\
5.35	0.744962477601931\\
5.45	0.742937477254825\\
5.55	0.740919780625773\\
5.65	0.738921227876835\\
5.75	0.736954561099739\\
5.85	0.735031248144029\\
5.95	0.73315982184389\\
6.05	0.731344882751639\\
6.15	0.729586790277606\\
6.25	0.727881968954847\\
6.35	0.72622368982775\\
6.45	0.724603151887944\\
6.55	0.723010682168515\\
6.65	0.721436890361526\\
6.75	0.719873647593133\\
6.85	0.718314801923146\\
6.95	0.716756588408368\\
7.05	0.715197733186704\\
7.15	0.713639284799467\\
7.25	0.712084229809002\\
7.35	0.710536962956784\\
7.45	0.709002685070524\\
7.55	0.7074867966637\\
7.65	0.705994344421724\\
7.75	0.704529563717926\\
7.85	0.703095544668799\\
7.95	0.701694033921699\\
8.05	0.700325370980316\\
8.15	0.698988547058625\\
8.25	0.697681366311799\\
8.35	0.696400684004567\\
8.45	0.695142693816228\\
8.55	0.693903236682071\\
8.65	0.692678105755258\\
8.75	0.691463325743889\\
8.85	0.69025538959911\\
8.95	0.689051440846107\\
9.05	0.687849395269387\\
9.15	0.686648000684952\\
9.25	0.685446837811119\\
9.35	0.684246268721207\\
9.45	0.683047342009644\\
9.55	0.681851665446273\\
9.65	0.680661257440358\\
9.75	0.679478388359084\\
9.85	0.678305421981977\\
9.95	0.677144666152418\\
};
\end{axis}
\end{tikzpicture}%
\begin{tikzpicture}

\begin{axis}[%
width=.4\textwidth,
height=2cm, 
scale only axis,
xmin=1,ylabel={$y$},xlabel={$x_1$},y label style={at={(-0.05,0.5)}},
xmax=11,
ymin=0,
ymax=3,
axis background/.style={fill=white}
]
\addplot [color=mycolor1, draw=none, mark=x, mark options={solid, mycolor1}, forget plot]
  table[row sep=crcr]{%
1	2.65\\
2	1.95\\
3	1.45\\
4	1.35\\
5	1.25\\
6	1.05\\
7	0.75\\
8	0.649999999999999\\
9	0.45\\
10	0.649999999999999\\
};
\addplot [color=mycolor2, forget plot]
  table[row sep=crcr]{%
1	2.45\\
2	2.08281825883856\\
3	1.72297166354556\\
4	1.43203695709684\\
5	1.21690498767597\\
6	1.05454638058171\\
7	0.880512997649619\\
8	0.689425568595503\\
9	0.533871365035334\\
10	0.414391332403024\\
11	0.391748675132592\\
};
\addplot [color=mycolor3, draw=none, mark=x, mark options={solid, mycolor3}, forget plot]
  table[row sep=crcr]{%
1	2.65\\
2	1.95\\
3	1.45\\
4	1.25\\
5	1.05\\
6	0.75\\
7	0.55\\
8	0.55\\
9	0.55\\
10	0.45\\
};
\addplot [color=mycolor4, forget plot]
  table[row sep=crcr]{%
1	2.45\\
2	2.0871113747558\\
3	1.7428081605892\\
4	1.43212362636695\\
5	1.19378301181628\\
6	0.955399219287496\\
7	0.7311315249561\\
8	0.546350199258822\\
9	0.439466169410851\\
10	0.347023167036659\\
11	0.282554458630428\\
};
\addplot [color=mycolor5, draw=none, mark=x, mark options={solid, mycolor5}, forget plot]
  table[row sep=crcr]{%
1	2.65\\
2	1.95\\
3	1.45\\
4	1.35\\
5	1.05\\
6	0.95\\
7	0.95\\
8	0.75\\
9	0.75\\
10	0.55\\
};
\addplot [color=mycolor6, forget plot]
  table[row sep=crcr]{%
1	2.45\\
2	2.08437242029748\\
3	1.72964250424555\\
4	1.43256266099907\\
5	1.21546678473743\\
6	0.992270428898916\\
7	0.839742570952465\\
8	0.72036722767742\\
9	0.600287536685836\\
10	0.50011691129157\\
11	0.379700889442169\\
};
\addplot [color=mycolor7, draw=none, mark=x, mark options={solid, mycolor7}, forget plot]
  table[row sep=crcr]{%
1	2.65\\
2	1.95\\
3	1.45\\
4	1.45\\
5	1.15\\
6	1.05\\
7	0.850000000000001\\
8	0.55\\
9	0.55\\
10	0.350000000000001\\
};
\addplot [color=mycolor1, forget plot]
  table[row sep=crcr]{%
1	2.45\\
2	2.08325216234967\\
3	1.72644664053003\\
4	1.44940682976872\\
5	1.26307576843685\\
6	1.04449841608698\\
7	0.871148580223421\\
8	0.710470267671952\\
9	0.530648338704551\\
10	0.398506689964389\\
11	0.27030802594444\\
};
\addplot [color=mycolor2, draw=none, mark=x, mark options={solid, mycolor2}, forget plot]
  table[row sep=crcr]{%
1	2.65\\
2	1.85\\
3	1.35\\
4	1.05\\
5	0.95\\
6	0.75\\
7	0.75\\
8	0.75\\
9	0.850000000000001\\
10	0.75\\
};
\addplot [color=mycolor3, forget plot]
  table[row sep=crcr]{%
1	2.45\\
2	2.0772947215596\\
3	1.68891743467978\\
4	1.3717482384539\\
5	1.10213105092041\\
6	0.89691729369732\\
7	0.716580065595192\\
8	0.597768286705844\\
9	0.523060979159613\\
10	0.487065777871116\\
11	0.405949067969639\\
};
\addplot [color=mycolor4, draw=none, mark=x, mark options={solid, mycolor4}, forget plot]
  table[row sep=crcr]{%
1	2.65\\
2	1.85\\
3	1.45\\
4	1.15\\
5	1.05\\
6	0.95\\
7	0.95\\
8	0.649999999999999\\
9	0.55\\
10	0.45\\
};
\addplot [color=mycolor5, forget plot]
  table[row sep=crcr]{%
1	2.45\\
2	2.07758538156955\\
3	1.69809006231344\\
4	1.41307036584877\\
5	1.13738862532075\\
6	0.952014365546549\\
7	0.797246278063067\\
8	0.688231149487747\\
9	0.524584574827064\\
10	0.410446498436523\\
11	0.315811946468197\\
};
\addplot [color=mycolor6, draw=none, mark=x, mark options={solid, mycolor6}, forget plot]
  table[row sep=crcr]{%
1	2.65\\
2	2.05\\
3	1.55\\
4	1.25\\
5	0.95\\
6	0.850000000000001\\
7	0.850000000000001\\
8	0.75\\
9	0.75\\
10	0.45\\
};
\addplot [color=mycolor7, forget plot]
  table[row sep=crcr]{%
1	2.45\\
2	2.08757422741894\\
3	1.7549242426729\\
4	1.47743988099213\\
5	1.19200132120817\\
6	0.959745423464698\\
7	0.780160877253685\\
8	0.665133984079329\\
9	0.550736642387298\\
10	0.463973836593935\\
11	0.330710595483775\\
};
\addplot [color=mycolor1, draw=none, mark=x, mark options={solid, mycolor1}, forget plot]
  table[row sep=crcr]{%
1	2.65\\
2	1.85\\
3	1.15\\
4	1.05\\
5	0.95\\
6	0.850000000000001\\
7	0.649999999999999\\
8	0.649999999999999\\
9	0.649999999999999\\
10	0.55\\
};
\addplot [color=mycolor2, forget plot]
  table[row sep=crcr]{%
1	2.45\\
2	2.07768139490263\\
3	1.68104374922527\\
4	1.32335770020432\\
5	1.07622543290764\\
6	0.879841144887135\\
7	0.70973102546008\\
8	0.547717930397934\\
9	0.470549027258106\\
10	0.398957098087294\\
11	0.326394933925277\\
};
\addplot [color=mycolor3, draw=none, mark=x, mark options={solid, mycolor3}, forget plot]
  table[row sep=crcr]{%
1	2.65\\
2	1.95\\
3	1.25\\
4	0.95\\
5	0.75\\
6	0.649999999999999\\
7	0.45\\
8	0.350000000000001\\
9	0.25\\
10	0.350000000000001\\
};
\addplot [color=mycolor4, forget plot]
  table[row sep=crcr]{%
1	2.45\\
2	2.08105094144378\\
3	1.69914927389061\\
4	1.33438418443958\\
5	1.05752257857596\\
6	0.818949754117422\\
7	0.615968525378736\\
8	0.446628460712029\\
9	0.314315876454544\\
10	0.185747938625319\\
11	0.153017814510057\\
};
\addplot [color=mycolor5, draw=none, mark=x, mark options={solid, mycolor5}, forget plot]
  table[row sep=crcr]{%
1	2.65\\
2	1.95\\
3	1.35\\
4	1.05\\
5	1.05\\
6	0.850000000000001\\
7	0.75\\
8	0.75\\
9	0.649999999999999\\
10	0.55\\
};
\addplot [color=mycolor6, forget plot]
  table[row sep=crcr]{%
1	2.45\\
2	2.08250274694372\\
3	1.71766155327182\\
4	1.39866695369378\\
5	1.12459444091449\\
6	0.939007741931164\\
7	0.765132051656333\\
8	0.615380404070769\\
9	0.5375845282806\\
10	0.442918007569914\\
11	0.351034337112777\\
};
\end{axis}
\end{tikzpicture}%
\caption{ On the left: $(\eps,\delta)$-robust satisfaction probability of $\eventually\always^{\le n_2}\{y\in [-2,2]\}$ with $\eps= 1.2266$ and $\delta=0.03$. On the right: simulation runs for the original model and the abstract model with the composed robust controller.}
\label{fig:simresult}\mbox{} \\[-.7cm]
\end{figure}
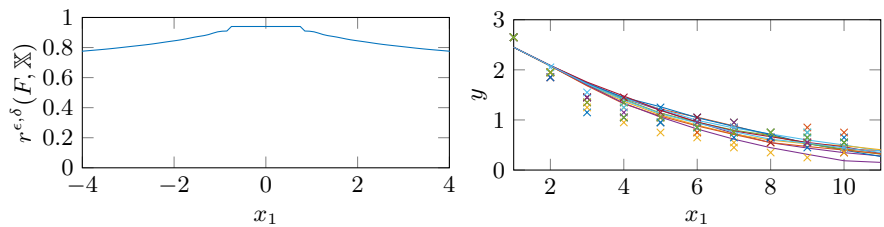

%% file: Appendix.tex
\section{Additional proofs}

\begin{proof}[\textbf{Proof of Theorem~\ref{thm:delreach}}]
  Consider any two models $\M_1,\M_2\in\mathcal M_\Y$ with $\M_1\preceq^\delta_0 \M_2$.
A Markov policy $\mu=(\mu_0,\ldots,\mu_N)$ for $\M_1$ can be refined to $\M_2$. Technically, this means that we first can write it as a control strategy for which we can do control refined as proven in \cite{haesaert2017verification}.
Denote the control strategy that refines the Markov policy as $\Ca_2$. Then the composed system $\Ca_2\times\M_2$ contains transitions over $\X_1\times \X_2$ with stochastic transition kernels defined as
\[\Wt(d\bar x_1\times d\bar x_2|\mu_k(x_1), x_1,x_2)\mbox{ for } k \in\{0,\ldots, N-1\}\]

We now need to show that $r^\mu(K_{\X_1},N)$ defined in \eqref{eq:prob1} is a lower bound for the  probability that target set $K$ is reached by $\Ca_2\times\M_2$ in $N$ time steps,
$\pcm{\Ca_2}{\M_2}(\lozenge^{\le N} K)$.
Onserve that the following holds
\begin{equation}
\label{eq:toRK}
\pcm{\Ca_2}{\M_2}\left[ \left((x_1, x_2)\in \rel \right)
   \mathcal U^{\leq N} \left(( x_1, x_2)\in \rel\wedge (x_1\in K_{\X_1}) \right)\right]\leq   \pcm{\Ca_2}{\M_2}\left[ \lozenge^{\leq N} (y_2\in K) \right],
\end{equation}
since $\left(( x_1, x_2)\in R\wedge (x_1\in K_{\X_1}) \right)$ implies $h_2(x_2)=h_1(x_1)$ and $h_1(x_1)\in K$.
The left-hand side of \eqref{eq:toRK} can be computed via backward recursion of reach-avoid property. For this the safe set (i.e., the complement of the avoid set) is $\rel$ and the reach set is $G:=(K_{\X_1}\times \X_2) \cap \rel$.
The value functions $\{V_k,\,k=0,\ldots,N\}$ in the backward recursion are initialised with  $V_N =0$ and computed as
\begin{align*}
V_k(x_1, x_2)&=\int\limits_{\X_1\times\X_2}\left[ \mathbf 1_{G}(\bar x_1)
+ \mathbf 1_{\rel \setminus G}(\bar x_1,\bar x_2)V_{k+1}(\bar x_1,\bar x_2)\right]\\
&\hspace{4.5cm}\times  \Wt(d \bar x_1\times d\bar x_2|\mu_k(x_1), x_1, x_2).
   \end{align*}
This can be also written down as
\begin{align*}
V_k(x_1, x_2)& =\int_{\rel}\left[ \mathbf 1_{(K_{\X_1}\times \X_2)}(\bar x_1)+ \mathbf 1_{(\X_1\setminus K_{\X_1})\times \X_2}(\bar x_1,\bar x_2)V_{k+1}(\bar x_1,\bar x_2)\right]\\
&\hspace{4.5cm}\times\Wt(d \bar x_1\times d\bar x_2|\mu_k(x_1), x_1, x_2).
\end{align*}
We now want to compute  a lower bound on
 $V_k(x_1,x_2)$ based on backwards computations over $ \M_1$. 
 The value functions $V^\delta_k: \X_1\rightarrow [0,1]$ are defined inductively as $V^\delta_k = \mathbf T_\delta^{\mu_k}(V^\delta_{k+1})$, which is
 \begin{equation*}
 V^\delta_k( x_1) := \Lim\left(\int_{\X_1}
 \left[\mathbf 1_{K_{\X_1}}(\bar x_1) +  \mathbf 1_{ \X_1\setminus K_{\X_1}}(\bar x_1)  V^{\delta}_{k+1}( \bar x_1)\right]
 \mathbb T(d\bar x_1|x_1,\mu_k(x_1))-\delta\right)
 \end{equation*}
and initialised with $V^\delta_k=0$.
With focus on the above two recursions we claim that if $V^\delta_{k+1}(x_1)$ is a lower bound for  $V_{k+1}(x_1,x_2)$ for all $(x_1,x_2)\in \rel$, then $V^\delta_{k}(x_1)$ is also a lower bound for  $V_{k}(x_1,x_2)$.
Once we prove this claim, by induction we get that $V_0^\delta(x_1)$ is a lower bound for $V_{k+1}(x_1,x_2)$. As a result, the $\delta$-robust probability can be computed as
\begin{equation*}
r^\mu(K_{\X_1},N):=\Lim\left(\int_{\X_1}[ \mathbf 1_{K_{\X_1}}(\bar x_1) +\mathbf 1_{ \X_1 \setminus K_{\X_1}}(\bar x_1)V_0^\delta(\bar x_1)] \pi(d  \bar x_1)-\delta\right).    
\end{equation*}

To prove the claim we need to define $\tilde V^\delta_k :\X_1\rightarrow [-\delta, 1]$ as
\begin{equation*}
\tilde V^\delta_k(x_1):=
 \int_{\X_1}
 \left[\mathbf 1_{K_{\X_1}}(\bar x_1) +  \mathbf 1_{ \X_1\setminus K_{\X_1}}(\bar x_1)  V^{\delta}_{k+1}( \bar x_1)\right]
 \mathbb T(d\bar x_1|x_1,\mu_k(x_1))-\delta,
 \end{equation*}
such that $V^\delta_k( x_1) = \Lim\left(\tilde V^\delta_k(x_1)\right)$.
For any $(x_1,x_2)\in \rel$, we have
\begin{align}
  \tilde V^\delta_{k}(x_1)+\delta& =\int\limits_{ \X_1}
  \left[\mathbf 1_{K_{\X_1}}(\bar x_1) +  \mathbf 1_{ \X_1 \setminus K_{\X_1}}(\bar x_1)  V^\delta_{k+1}(\bar x_1)\right]
  \mathbb T(d \bar x_1|x_1,\mu_k(x_1))\notag\\
  & = \int\limits_{\X_1\times\X_2}\left[ \mathbf 1_{(K_{\X_1}\times \X_2)}(\bar x_1,\bar x_2)+ \mathbf 1_{(\X_1\setminus K_{\X_1})\times \X_2}(\bar x_1,\bar x_2) V^\delta_{k+1}(\bar x_1)\right]\label{eq:relation_reas}\\
  &\hspace{5cm}\times\Wt(d \bar x_1\times d\bar x_2|\mu_k(x_1), x_1, x_2)\notag\\
  &\leq \int\limits_{\rel}\left[ \mathbf 1_{(K_{\X_1}\times \X_2)}(\bar x_1,\bar x_2)+ \mathbf 1_{(\X_1\setminus K_{\X_1})\times \X_2}(\bar x_1,\bar x_2) V_{k+1}(\bar x_1,\bar x_2)\right]\notag\\
  &\hspace{5cm}\times\Wt(d \bar x_1\times d\bar x_2|\mu_k( x_1), x_1, x_2)\notag\\
  & \hspace{2cm}+\int\limits_{(\X_1\times \X_2)\setminus \rel} \Wt(d \bar x_1\times d\bar x_2|\mu_k( x_1), x_1, x_2)\label{eq:bound}\\
  &\leq \int\limits_{\rel}\left[ \mathbf 1_{(K_{\X_1}\times \X_2)}(\bar x_1,\bar x_2)+ \mathbf 1_{(\X_1\setminus K_{\X_1})\times \X_2}(\bar x_1,\bar x_2) V_{k+1}( \bar x_1,\bar x_2)\right]\notag\\
  &\hspace{4.5cm}\times\Wt(d \bar x_1\times d\bar x_2|\mu_k( x_1), x_1, x_2) +\delta.
  \notag\end{align}
  Note that \eqref{eq:relation_reas} follows from the fact that $(x_1,x_2)\in\rel$ and \eqref{eq:bound} holds since $V_{k+1}^\delta(\bar x_1)$ is upper bounded by $1$ and is lower bounded by $V_{k+1}^{\delta}(\bar x_1,\bar x_2)$ over $\rel$.
Thus by the definition of $V_k$, we have also that $V_k(x_1,x_2)\le \tilde V_k^\delta(x_1)$ for all $(x_1,x_2)\in \rel$.
Since $V_k$ is also lower bounded by $0$, we have that $V_k(x_1,x_2)\le\Lim\left(\tilde V_k^\delta(x_1)\right) = V_k^\delta(x_1)$, which completes the proof.


\end{proof}

\begin{proof}[\textbf{Proof of Theorem~\ref{thm:maxprob}}]
This proof upper-bounds the maximal reachability probability and follows an analogue path to the proof of Theorem \ref{thm:delreach}.
	  Consider any two models $\M_1,\M_2\in\mathcal M_\Y$ with $\M_1\succeq^\delta_\eps \M_2$.
A Markov policy $\mu=(\mu_0,\ldots,\mu_N)$ for $\M_2$ can be refined to $\M_1$ such that the composed system $\Ca_1\times\M_1$ contains transitions over $\X_1\times \X_2$ with stochastic transition kernels defined as
\[\Wt(d\bar x_1\times d\bar x_2|\mu_k(x_2), x_1,x_2)\mbox{ for } k \in\{0,\ldots, N-1\}.\]
We now need to show that  $\pcm{\mu_2}{\M_2}(\lozenge^{\le N} K)$ has an upper bound computed as given in Theorem \ref{thm:maxprob}.
Remark that $\lozenge^{\leq N} (y_2\in K)$ implies the specification
\[\psi:= \left[(x_1, x_2)\in \rel \right]
   \mathcal U^{\leq N} \left[( x_1, x_2)\in \bar \rel\vee (x_1\in K_{\X_1}^{-\eps}) \right]
\]
with $\bar \rel:=\X_1\times\X_2\setminus \rel$. Hence it follows that
\begin{align}\label{eq:toRK2}
  \pcm{\mu_2}{\M_2}\left[ \lozenge^{\leq N} (y_2\in K) \right]\leq \pcm{\Ca_1}{\M_1}\left[\psi\right].
\end{align}

The right-hand side of \eqref{eq:toRK2} can be computed via backward recursion of reach-avoid property. For this the safe set (i.e., the complement of the avoid set) is $\rel$ and the reach set is $((K_{\X_1}\times \X_2) \cap \rel) \cup \bar \rel $.
The value functions $V_k$ in the backward recursion are initialised with $V_N =0$ and computed as
\begin{align}
&V_k(x_1, x_2)=\int\limits_{\X_1\times\X_2}\left[ \mathbf 1_{(K_{\X_1}^{-\eps}\times \X_2)\cap\rel}(\bar x_1,\bar x_2)+\mathbf 1_{\bar\rel}(\bar x_1,\bar x_2)\right.\nonumber\\
&\left.\quad\qquad+ \mathbf 1_{ \rel \setminus ( K_{\X_1}^{-\eps}\times\X_2)}(\bar x_1,\bar x_2)V_{k+1}( \bar x_1,\bar x_2)\right]  \Wt(d \bar x_1\times d\bar x_2|\mu_k(x_2), x_1, x_2).
\label{eq:ap:Vkproof2}
\end{align}
  We now want to compute an upper bound on $V_k(x_1,x_2)$ based on backward recursion over $\M_1$.  Let $V^{-\delta}_k: \X_1\rightarrow [0,1]$ be defined inductively as
 \[ V^{-\delta}_k( x_1) :=\sup_{\bar\mu} \Lim\left(\int_{ \X_1}
 \left[\mathbf 1_{K_{\X_1}}(\bar x_1) +  \mathbf 1_{ \X_1 \setminus K_{\X_1}}(\bar x_1)  V^{-\delta}_{k+1}(\bar x_1)\right]
 \mathbb T(d\bar x_1|\bar\mu_k(x_1), x_1)+\delta\right)\]
and initialised with $V^{-\delta}_N=0$.
We claim that if $V^{-\delta}_{k+1}(x_1)\ge V_{k+1}(x_1,x_2)$ for all $(x_1,x_2)\in \rel$, then $V^{-\delta}_{k}(x_1)\ge V_{k}(x_1,x_2)$.
Thus by induction we get $V^{-\delta}_0 (x_1)\ge V_0(x_1,x_2)$. By repeating the same argument for initial measure $\pi$, we get Theorem~\ref{thm:maxprob}.

In order to prove the above claim we need to define value functions
$\tilde V^\delta_k :\X_1\rightarrow [-\delta, 1]$ as
 \[ \tilde V^{-\delta}_k(x_1):=\sup_{\bar\mu}\left(\int_{ \X_1}
 \left[\mathbf 1_{K_{\X_1}}(\bar x_1) +  \mathbf 1_{ \X_1 \setminus K_{\X_1}}(\bar x_1)  V^{-\delta}_{k+1}(\bar x_1)\right]
 \mathbb T(d\bar x_1|\bar\mu_k(x_1), x_1)\right)+\delta,\]
 such that $V^{-\delta}_k(x_1) = \Lim\left(\tilde V^{-\delta}_k(x_1)\right)$ due to the fact that $\Lim$ and $\sup$ are interchangeable here.
For any $(x_1,x_2)\in \rel$ we have
\begin{align*}
V_k(x_1,x_2)&\leq \int\limits_{\rel}\left[ \mathbf 1_{(K_{\X_1}^{-\eps}\times \X_2)\cap\rel}(\bar x_1,\bar x_2)+ \mathbf 1_{\rel \setminus (K_{\X_1}^{-\eps}\times\X_2)}(\bar x_1,\bar x_2)V_{k+1}( \bar x_1,\bar x_2)\right]\\&\quad\qquad\times  \Wt(d \bar x_1\times d\bar x_2|\mu_k( x_2), x_1, x_2) + \delta\\
&\leq \int\limits_{\rel}\left[ \mathbf 1_{(K_{\X_1}^{-\eps}\times \X_2)}(\bar x_1,\bar x_2)+ \mathbf 1_{(\X_1\times \X_2)\setminus (K_{\X_1}^{-\eps}\times\X_2)}(\bar x_1,\bar x_2)V_{k+1}^{-\delta}( \bar x_1)\right]\\&\quad\qquad\times  \Wt(d \bar x_1\times d\bar x_2|\mu_k( x_2), x_1, x_2) + \delta\\
&\leq \int\limits_{\X_1\times\X_2}\left[ \mathbf 1_{(K_{\X_1}^{-\eps}\times \X_2) }(\bar x_1,\bar x_2)+ \mathbf 1_{(\X_1\times \X_2)\setminus (K_{\X_1}^{-\eps}\times\X_2)}(\bar x_1,\bar x_2)V_{k+1}^{-\delta}(\bar x_1)\right]\\&\quad\qquad\times  \Wt(d \bar x_1\times d\bar x_2|\mu_k( x_2), x_1, x_2) + \delta\\
&\leq \sup_{\bar \mu}\int\limits_{\X_1}\left[ \mathbf 1_{K_{\X_1}^{-\eps} }(\bar x_1)+ \mathbf 1_{\X_1 \setminus K_{\X_1}^{-\eps}}(\bar x_1)V_{k+1}^{-\delta}(\bar x_1)\right] \mathbb T(d\bar x_1|\bar\mu(x_1),x_1) + \delta\\
& =\tilde V^{-\delta}_k(x_1).
   \end{align*}
  Since $V_k(x_1,x_2):\X_1\times \X_2 \rightarrow[0,1]$ it holds that if 
   $V_k(x_1,x_2)\leq \tilde V^{-\delta}_k(x_1)$ then also $V_k(x_1,x_2)\leq  \Lim\left(\tilde V^{-\delta}_k(x_1)\right) = V^{-\delta}_k(x_1)$. 
   This completes the proof.

\end{proof}

\begin{proof} [\textbf{Proof of Theorem~\ref{thm:DFA_product}}]
	Since $\M_1\preceq_0^\delta\M_2$, according to Definition~\ref{def:apbsim} there exists an interface function $\InF$,
	a relation $\rel\subseteq \X_1\times \X_2$, and a Borel measurable stochastic kernel $\Wt(\,\cdot\,{\mid} u_1,x_1,x_2)$
	such that
	\begin{enumerate}
		\item $\forall (x_1,x_2)\in \rel$, $ h_1(x_1)=h_2(x_2)$;
		\item $\forall (x_1,x_2)\in \rel$, $\forall {u_1}\in \A_1,$
		\(\mathbb T_1(\cdot| x_1,u_1)\ \bar \rel_\delta\  \mathbb T_2(\cdot| x_2,\InF(u_1,x_1,x_2)),\) with lifted probability measure $\Wt(\,\cdot\,{\mid} u_1,x_1,x_2)$;
		\item $\pi_1\bar\rel_\delta \pi_2$.
	\end{enumerate}
	Indicate the product gMDPs by $\M_i\otimes\mathcal A = (\bar{\X}_i,\bar\pi_i,\bar{\mathbb T}_i,\A,\bar h_i,\Y)$. According to Definition~\ref{def:product} we have for any $i=1,2$,
	$\bar{\X}_i = \X_i\times Q$, $\bar\pi_i(dx_i,q) = \pi_i(dx_i)\cdot1(q = q_0)$, $\bar h_i(x_i,q) = h_i(x_i)$ for any $(x_i,q)\in\bar{\X}_i$, and
	\begin{equation*}
	\bar{\mathbb T}_i(A_i\times\{q'\}|x_i,q,u_i) = \int_{\tilde x_i\in A_i}\mathbf 1(q' = \trans(q,\mathsf L(h_i(\tilde x_i))))\cdot\mathbb T_i(d\tilde x_i|x_i,u_i).
	\end{equation*}
	In order to prove the theorem we construct the relation $\rel^{\mathsf p}$, the interface function $\InF^{\mathsf p}$, and the lifted measure $\Wt^{\mathsf p}$ based on $\rel,\InF,\Wt$.
	\begin{enumerate}
		\item  Define $\rel^{\mathsf p}\subseteq\bar\X_1\times\bar\X_2$ with $(x_1,q_1)\rel^{\mathsf p}(x_2,q_2)$ iff $x_1\rel x_2$ and $q_1=q_2$.
		Select any $\bar x_1 = (x_1,q_1)\in\bar\X_2$ and $\bar x_2 = (x_2,q_2)\in\bar\X_2$. Then
		\begin{equation*}
		\bar h_1(\bar x_1) = \bar h_1(x_1,q_1) = h_1(x_1) \text{ and } \bar h_2(\bar x_2) = \bar h_2(x_2,q_2) = h_2(x_2).
		\end{equation*}
		Thus
		\begin{equation*}
		(\bar x_1,\bar x_2)\in\rel^{\mathsf p}\Rightarrow (x_1,x_2)\in\rel\Rightarrow h_1(x_1) = h_2(x_2)\Rightarrow \bar h_1(\bar x_1) = \bar h_2(\bar x_2).
		\end{equation*}
		
		\item Since $\pi_1\bar\rel_\delta \pi_2$, there exists a lifted measure $\mathbb W_{\textsf{Init}}$ such that
		$\mathbb W_{\textsf{Init}}(\rel)\ge  1-\delta$,
		$\mathbb W_{\textsf{Init}}(A_1\times \X_2)=\pi_1(A_1)$ for all $A_1\in \mathcal{B}(\X_1)$,
		and $\mathbb W_{\textsf{Init}}(\X_1\times A_2)=\pi_2(A_2)$ for all $A_2\in \mathcal{B}(\X_2)$.
		Define the probability space $\left(\bar\X_1\times\bar\X_2,\mathcal B(\bar\X_1\times\bar\X_2),\bar{\mathbb W}_{\textsf{Init}}\right)$ with the property that for all $A_1\in\mathcal B(\X_1)$, $A_2\in\mathcal B(\X_2)$, and $q_1,q_2\in Q$
		\begin{equation*}
			\bar{\mathbb W}_{\textsf{Init}}(\{q_1\}\times A_1\times\{q_2\}\times A_2) := \mathbb W_{\textsf{Init}}(A_1\times A_2)1(q_1 = q_0)1(q_2=q_0),
		\end{equation*}
		where $q_0$ is the initial state of the automaton $\mathcal A$. In words, $\bar{\mathbb W}_{\textsf{Init}}$ assigns probabilities (which are the same as $\mathbb W_{\textsf{Init}}$) to Borel measurable subsets of $\bar\X_1\times\bar\X_2$ if and only if the discrete modes are equal to $q_0$. For this particular lifted measure, we have
		\begin{itemize}
			\item $\bar{\mathbb W}_{\textsf{Init}}(\{q_1\}\times A_1\times \bar\X_2)=\mathbb W_{\textsf{Init}}(A_1\times \X_2)1(q_1 = q_0) = \pi_1(A_1)1(q_1 = q_0) = \bar\pi_1(\{q_1\}\times A_1).$
			\item $\bar{\mathbb W}_{\textsf{Init}}(\bar\X_1\times \{q_2\}\times A_2)=\mathbb W_{\textsf{Init}}(\X_1\times A_2)1(q_2 = q_0) = \pi_2(A_2)1(q_2 = q_0) = \bar\pi_2(\{q_2\}\times A_2).$
			\item $\bar{\mathbb W}_{\textsf{Init}}(\rel^{\mathsf p}) = \sum_{q_1,q_2}\bar{\mathbb W}_{\textsf{Init}}((x_1,x_2)\in\rel\,\wedge\,q_1 = q_2) = \mathbb W_{\textsf{Init}}(\rel) \ge 1-\delta$.
		\end{itemize}
		Therefore $\bar\pi_1\bar\rel^{\mathsf p}_\delta\bar\pi_2$ with lifted measure $\bar{\mathbb W}_{\textsf{Init}}$.
		
		\item We know that for all $(x_1,x_2)\in \rel$ and ${u_1}\in \A_1$,
		\(\mathbb T_1(\cdot| x_1,u_1)\ \bar \rel\  \mathbb T_2(\cdot| x_2,\InF(u_1,x_1,x_2)),\) with lifted probability measure $\Wt(\,\cdot\,{\mid} u_1,x_1,x_2)$.
		Define the new measure $\Wtp(\,\cdot\,{\mid} u_1,\bar x_1,\bar x_2)$ on $\bar\X_1\times\bar\X_2$ given $\A\times\bar\X_1\times\bar\X_2$ with
		\begin{equation}
		\label{eq:lifted_meas_prod}
			\begin{split}
			&\Wtp(\{q'_1\}\times A_1\times\{q'_2\}\times A_2\,|\, u_1,q_1,x_1,q_2,x_2) :=
			\int_{\tilde x_1\in A_1}\int_{\tilde x_2\in A_2}\ldots\\
			&\mathbf 1(q'_1 = \trans(q_1,\mathsf L(h_1(\tilde x_1))))
			\cdot \mathbf 1(q'_2 = \trans(q_2,\mathsf L(h_2(\tilde x_2))))\cdot
			\Wt(d\tilde x_1\times d\tilde x_2{\mid} u_1,x_1,x_2).
			\end{split}
		\end{equation}
		In words, $\Wtp$ assigns probabilities to Borel measurable subsets of $\bar\X_1\times\bar\X_2$ which are the same probabilities as in $\Wt$ and evolves the discrete mode of the two gMDP according to the automaton $\mathcal A$. For this particular lifted measure $\Wtp$, we have for any $\bar x_1 = (x_1,q_1)\in\X_1$ and $\bar x_2 = (x_2,q_2)\in\X_2$ with $\bar x_1\rel^{\mathsf p}\bar x_2$ and any
		$u_1\in\A_1 $:
		\begin{align*}
		\Wtp(\{q'_1\}&\times A_1\times \bar\X_2\,|\, u_1,q_1,x_1,q_2,x_2)
		= \sum_{q_2'\in Q}\int_{\tilde x_2\in \X_2}\int_{\tilde x_1\in A_1}\ldots\\
		& = \int_{\tilde x_1\in A_1} \mathbf 1(q'_1 = \trans(q_1,\mathsf L(h_1(\tilde x_1))))
			\int_{\tilde x_2\in \X_2}\Wt(d\tilde x_1\times d\tilde x_2{\mid} u_1,x_1,x_2)\\
		&=\int_{\tilde x_1\in A_1} \mathbf 1(q'_1 = \trans(q_1,\mathsf L(h_1(\tilde x_1))))
			\Wt(d\tilde x_1\times \X_2{\mid} u_1,x_1,x_2)\\
			&=\int_{\tilde x_1\in A_1} \mathbf 1(q'_1 = \trans(q_1,\mathsf L(h_1(\tilde x_1))))
			\mathbb T_1(d\tilde x_1|x_1,u_1)\\
		&= \bar{\mathbb T}_1(A_1\times\{q'_1\}|x_1,q_1,u_1),
		\end{align*}
		and
		\begin{align*}
		\Wtp(\bar\X_1&\times \{q'_2\}\times A_2\,|\, u_1,q_1,x_1,q_2,x_2)
		=\sum_{q_1'\in Q}\int_{\tilde x_1\in \X_1}\int_{\tilde x_2\in A_2}\ldots\\
		& = \int_{\tilde x_2\in A_2} \mathbf 1(q'_2 = \trans(q_2,\mathsf L(h_2(\tilde x_2))))
			\int_{\tilde x_1\in \X_1}\Wt(d\tilde x_1\times d\tilde x_2{\mid} u_1,x_1,x_2)\\
		& = \int_{\tilde x_2\in A_2} \mathbf 1(q'_2 = \trans(q_2,\mathsf L(h_2(\tilde x_2))))
		\Wt(\X_1\times d\tilde x_2{\mid} u_1,x_1,x_2)\\
	    & = \int_{\tilde x_2\in A_2} \mathbf 1(q'_2 = \trans(q_2,\mathsf L(h_2(\tilde x_2))))
		\mathbb T_2(d\tilde x_2|x_2,\InF(u_1,x_1,x_2))\\
		&= \bar{\mathbb T}_2(A_2\times\{q'_2\}|x_2,q_2,\InF(u_1,x_1,x_2)).
		\end{align*}
		Take any $(x_1,q_1)\rel^{\mathsf p}(x_2,q_2)$ which implies $q_1 = q_2$ and $x_1\rel x_2$, hence $h_1(x_1) = h_2(x_2)$. If we also assume $(\tilde x_1,q'_1)\rel^{\mathsf p}(\tilde x_2,q'_2)$, then
		\begin{equation*}
		\trans(q_1,\mathsf L(h_1(\tilde x_1))) = \trans(q_2,\mathsf L(h_2(\tilde x_2))),
		\end{equation*}
		and we get
		\begin{align*}
		\Wtp(\rel^{\mathsf p}\,|\, u_1,q_1,&x_1,q_2,x_2) = \int_{(\tilde x_1,\tilde x_2)\in\rel}\Wt(d\tilde x_1\times d\tilde x_2{\mid} u_1,x_1,x_2)\\
		& \cdot \sum_{q'_1,q'_2\in Q}\mathbf 1(q'_1 = \trans(q_1,\mathsf L(h_1(\tilde x_1))))\cdot \mathbf 1(q'_2 = \trans(q_2,\mathsf L(h_2(\tilde x_2)))).
		\end{align*}
		The above sum is equal to one due to $q_1 = q_2$, $(\tilde x_1,\tilde x_2)\in\rel$, and the DFA being deterministic. Then
		\begin{align*}
		\Wtp(\rel^{\mathsf p}\,|\, u_1,q_1,&x_1,q_2,x_2) = \int_{\rel}\Wt(d\tilde x_1\times d\tilde x_2{\mid} u_1,x_1,x_2) \ge 1-\delta.
		\end{align*}
		We have shown that
		\begin{equation*}
		\bar{\mathbb T}_1(A_1\times\{q'_1\}|x_1,q_1,u_1)\bar\rel^{\mathsf p}_\delta\bar{\mathbb T}_2(A_2\times\{q'_2\}|x_2,q_2,\InF(u_1,x_1,x_2))
		\end{equation*}
		with lifted measure $\Wtp$ defined in \eqref{eq:lifted_meas_prod} and the same interface function $\InF$.
	\end{enumerate}
\end{proof}

%% file: main.bbl
\begin{thebibliography}{10}
	
	\bibitem{Abate2011}
	A.~Abate.
	\newblock Approximation metrics based on probabilistic bisimulations for
	general state-space {M}arkov processes: a survey.
	\newblock {\em ENTCS}, 297:3--25, 2013.
	
	\bibitem{AbateQuanti}
	A.~Abate, J.-p. Katoen, and A.~Mereacre.
	\newblock {Quantitative Automata Model Checking of Autonomous Stochastic Hybrid
		Systems}.
	\newblock In {\em Proc. 14th ACM Int. Conf. Hybrid Syst. Comput. Control},
	pages 83--92, 2011.
	
	\bibitem{Abate1}
	A.~Abate, M.~Prandini, J.~Lygeros, and S.~Sastry.
	\newblock {Probabilistic Reachability and Safety for Controlled Discrete Time
		Stochastic Hybrid Systems}.
	\newblock {\em Automatica}, 44(11):2724--2734, 2008.
	
	\bibitem{Bible}
	D.~Bertsekas and S.~E. Shreve.
	\newblock {\em {Stochastic Optimal control : The discrete time case}}.
	\newblock Athena Scientific, 1996.
	
	\bibitem{bogachev2007measure}
	V.~I. Bogachev.
	\newblock {\em Measure theory}.
	\newblock Springer Science \& Business Media, 2007.
	
	\bibitem{Boyd2004}
	S.~Boyd and L.~Vandenberghe.
	\newblock {\em {Convex Optimization}}.
	\newblock CUP, Cambridge, 2004.
	
	\bibitem{Desharnais2008}
	J.~Desharnais, F.~Laviolette, and M.~Tracol.
	\newblock Approximate analysis of probabilistic processes: Logic, simulation
	and games.
	\newblock {\em Conf. on Quantitative Evaluation of Systems}, pages 264--273,
	Sept. 2008.
	
	\bibitem{cDAK12}
	A.~D'Innocenzo, A.~Abate, and J.-P. Katoen.
	\newblock Robust {PCTL} model checking.
	\newblock In {\em Proceedings of the 15th ACM international conference on
		Hybrid Systems: computation and control}, pages 275--285, 2012.
	
	\bibitem{SAID}
	S.~{Esmaeil Zadeh Soudjani} and A.~Abate.
	\newblock Adaptive and sequential gridding procedures for the abstraction and
	verification of stochastic processes.
	\newblock {\em SIAM Journal on Applied Dynamical Systems}, 12(2):921--956,
	2013.
	
	\bibitem{FAUST13}
	S.~{Esmaeil Zadeh Soudjani}, C.~Gevaerts, and A.~Abate.
	\newblock {FAUST $^2$: Formal Abstractions of Uncountable-STate STochastic
		Processes}.
	\newblock In {\em Tools and Algorithms for the Construction and Analysis of
		Systems (TACAS)}, Lecture Notes in Computer Science, pages 272--286. Springer
	Berlin Heidelberg, 2015.
	
	\bibitem{Girard2009}
	A.~Girard and G.~J. Pappas.
	\newblock {Hierarchical control system design using approximate simulation}.
	\newblock {\em Automatica}, 45(2):566--571, 2009.
	
	\bibitem{DBLP:conf/qest/HaesaertAH16}
	S.~Haesaert, A.~Abate, and P.~M.~J. {Van den Hof}.
	\newblock Verification of general markov decision processes by approximate
	similarity relations and policy refinement.
	\newblock In {\em 13th International Conference on Quantitative Evaluation of
		Systems, {QEST} 2016, Quebec City, Canada, August 23-25}, pages 227--243,
	2016.
	
	\bibitem{haesaert2017verification}
	S.~Haesaert, S.~{Esmaeil Zadeh Soudjani}, and A.~Abate.
	\newblock Verification of general markov decision processes by approximate
	similarity relations and policy refinement.
	\newblock {\em SIAM Journal on Control and Optimization}, 55(4):2333--2367,
	2017.
	
	\bibitem{tech_report_TACAS}
	S.~Haesaert, S.~Soudjani, and A.~Abate.
	\newblock {Temporal logic control of general Markov decision processes by
		approximate policy refinement}.
	\newblock Technical report, 10 2017.
	
	\bibitem{hll1996}
	O.~Hern{\'a}ndez-Lerma and J.~B. Lasserre.
	\newblock {\em Discrete-time {M}arkov control processes}, volume~30 of {\em
		Applications of Mathematics (New York)}.
	\newblock Springer Verlag, 1996.
	
	\bibitem{KupfermanVardi2001}
	O.~Kupferman and M.~Y. Vardi.
	\newblock Model checking of safety properties.
	\newblock {\em Formal Methods in System Design}, pages 291--314, 2001.
	
	\bibitem{KNP11}
	M.~Kwiatkowska, G.~Norman, and D.~Parker.
	\newblock {PRISM} 4.0: Verification of probabilistic real-time systems.
	\newblock In G.~Gopalakrishnan and S.~Qadeer, editors, {\em Proc. 23rd
		International Conference on Computer Aided Verification (CAV'11)}, volume
	6806 of {\em LNCS}, pages 585--591. Springer, 2011.
	
	\bibitem{larsen1991bisimulation}
	K.~G. Larsen and A.~Skou.
	\newblock Bisimulation through probabilistic testing.
	\newblock {\em Information and Computation}, 94(1):1--28, 1991.
	
	\bibitem{LSMZ17}
	A.~{Lavaei}, S.~{Esmaeil Zadeh Soudjani}, R.~{Majumdar}, and M.~{Zamani}.
	\newblock {Compositional Abstractions of Interconnected Discrete-Time
		Stochastic Control Systems}.
	\newblock {\em ArXiv e-prints}, Sept. 2017.
	
	\bibitem{mt1993}
	S.~P. Meyn and R.~L. Tweedie.
	\newblock {\em Markov chains and stochastic stability}.
	\newblock Communications and Control Engineering Series. Springer-Verlag London
	Ltd., 1993.
	
	\bibitem{safonov1989schur}
	M.~G. Safonov and R.~Chiang.
	\newblock A {Schur} method for balanced-truncation model reduction.
	\newblock {\em IEEE Transactions on Automatic Control}, 34(7):729--733, 1989.
	
	\bibitem{tmka2013}
	I.~Tkachev, A.~Mereacre, J.~Katoen, and A.~Abate.
	\newblock Quantitative automata-based controller synthesis for non-autonomous
	stochastic hybrid systems.
	\newblock In {\em HSCC}, pages 293--302, 2013.
	
\end{thebibliography}
